\documentclass[number,preprint,3p]{elsarticle}

\usepackage{nicefrac}
\usepackage{color}
\usepackage{graphicx}
\usepackage{subcaption}
\usepackage{algorithm}
\usepackage{bm}
\usepackage[colorlinks]{hyperref}
\usepackage{amssymb}
\usepackage{amsthm}
\usepackage{amsmath}
\usepackage{amssymb}
\usepackage{mathtools}
\usepackage{dsfont}
\usepackage{booktabs}
\usepackage{url}
\usepackage{scrextend}
\usepackage{epstopdf}
\usepackage{float}
\usepackage{tcolorbox}
\usepackage{tablefootnote}
\usepackage[perpage]{footmisc}
\usepackage{bbm}
\usepackage{lineno}
\usepackage{soul}

\def\defi{\coloneqq}
\def\tr{^\intercal}

\newcommand{\Prob}[1]{\mathbb{P}\left(#1\right)}
\newcommand{\E}[1]{\mathbb{E}\left[#1\right]}

\newcommand{\vect}[1]{\boldsymbol{#1}}


\makeatletter
\gdef\urlauthor#1#2{\g@addto@macro\@elsuads{\let\corref\@gobble%
     \def\@@tmp{#1}\raggedright\eadsep
     {\ttfamily\url{\expandafter\strip@prefix\meaning\@@tmp}}\space(#2)%
     \def\eadsep{\unskip,\space}}%
}
\gdef\emailauthor#1#2{\stepcounter{ead}%
     \g@addto@macro\@elseads{\raggedright%
      \let\corref\@gobble\def\@@tmp{#1}%
      \eadsep{\ttfamily\href{mailto:\expandafter\strip@prefix\meaning\@@tmp}{\expandafter\strip@prefix\meaning\@@tmp}}
      (#2)\def\eadsep{\unskip,\space}}%
}
\makeatother
\biboptions{numbers,sort&compress,square}
\journal{arXiv}

\begin{document}
\begin{frontmatter}
\renewcommand{\thefootnote}{\fnsymbol{footnote}}
\title{Long-range Ising model for regional-scale seismic risk analysis}
\author[1]{Sebin Oh}
\author[1]{Sangri Yi}
\author[1]{Ziqi Wang\corref{cor1}}
\ead{ziqiwang@berkeley.edu}  
\cortext[cor1]{Corresponding author}
\address[1]{Department of Civil and Environmental Engineering, University of California, Berkeley, United States}
\begin{abstract}
        \noindent
        This study introduces the long-range Ising model from statistical mechanics to the Performance-Based Earthquake Engineering (PBEE) framework for regional seismic damage analysis. The application of the PBEE framework at a regional scale involves estimating the damage states of numerous structures, typically performed using fragility function-based stochastic simulations. However, these simulations often assume conditional independence or employ simplistic dependency models among the damage states of structures, leading to significant misrepresentation of regional risk. The Ising model addresses this issue by converting the available information on binary damage states (safe or failure) into a joint probability mass function, leveraging the principle of maximum entropy. The Ising model offers two main benefits: (1) it requires only the first- and second-order cross-moments, enabling seamless integration with the existing PBEE framework, and (2) it provides meaningful physical interpretations of the model parameters, facilitating the uncovering of insights not apparent from data. To demonstrate the proposed method, we applied the Ising model to $156$ buildings in Antakya, Turkey, using post-hazard damage evaluation data, and to $182$ buildings in Pacific Heights, San Francisco, using simulated data from the Regional Resilience Determination (R2D) tool. In both instances, the Ising model accurately reproduces the provided information and generates meaningful insights into regional damage. The study also investigates the change in Ising model parameters under varying earthquake magnitudes, along with the mean-field approximation, further facilitating the applicability of the proposed approach.
\end{abstract}

\begin{keyword}
Ising model \sep performance-based earthquake engineering \sep regional seismic risk analysis \sep statistical mechanics
    
\end{keyword}
    
\end{frontmatter}

\renewcommand{\thefootnote}{\fnsymbol{footnote}}
    
\section{Introduction}		
\noindent 
The Performance-Based Earthquake Engineering (PBEE) framework has been extensively developed and utilized for over 20 years, providing a systematic method for risk-informed seismic design and post-hazard decision-making \cite{Moehle2004AEngineering, Porter2003AnMethodology}. To incorporate the various uncertainties that propagate from ground motions to the seismic performance of a structure, the PBEE framework considers four groups of random variables: (1) ground motion intensity measure (IM), (2) engineering demand parameter (EDP), (3) damage measure or damage state (DS), and (4) loss or decision variable (DV). {\color{black} Following the convention of probability theory, we use uppercase letters to represent random variables (e.g., DS) and lowercase letters to represent outcomes of random variables (e.g., ds).} Given an earthquake scenario, IM is typically described as a probability density function $f(im)$, or as a mean annual probability of exceedance $\lambda(im)$, assuming a Poisson process model for the temporal occurrence of earthquakes. This intensity measure model is followed by the conditional distribution of EDP given IM, denoted by $f(edp|im)$. Subsequently, the conditional distribution of DS given EDP, $f(ds|edp)$, is established, which can also be expressed as a complementary cumulative distribution function  $G(ds|edp)$. Finally, $f(ds|edp)$ is followed by the distribution of DV given DS, denoted by $f(dv|ds)$. To obtain the exceedance probability $\lambda(dv)$ of a decision variable, the probability density/mass functions $f(\cdot)$ are often transformed to complementary cumulative distribution functions $G(\cdot)$, yielding the integral:
\begin{equation} \label{Eq:PBEE}
    {\color{black}{\lambda(dv)=\int_{ds}\int_{edp}\int_{im}G(dv|ds)|d\,G(ds|edp)||d\,G(edp|im)||d\,\lambda(im)|\,.}}
\end{equation}
Building on Eq.~\eqref{Eq:PBEE}, the PBEE framework has been applied to a wide variety of structures, including braced steel frames \cite{Qiu2017Performance-basedBraces}, reinforced concrete moment-frame buildings \cite{Zhang2018ASafety,Haselton2007AssessingBuildings}, and highway bridges \cite{Li2020Long-termHazards}. These applications aim at individual structures, reflecting the original scope of the PBEE framework \cite{Heresi2023RPBEE:Scale}.

Recently, the application of the PBEE framework has been actively extended to a regional scale \cite{Heresi2023RPBEE:Scale,DeRisi2018Scenario-BasedScale,Hulsey2022High-resolutionCordons,Zhang2023Regional-scaleIstanbul}. In this context, Eq.~\eqref{Eq:PBEE} is typically solved using Monte Carlo methods that simulate regional responses to hazards, often referred to as regional seismic simulations. This simulation approach is favored because a system-level decision variable may depend on the damage states of all components, rendering Eq.~\eqref{Eq:PBEE} high-dimensional. Similar to the component-level applications, a regional-scale PBEE analysis can be decomposed into four modules: (1) IM generation, (2) EDP estimation, (3) DS evaluation, and (4) DV assessment. For each of these modules, physics-based or statistical methods/models can be employed, offering a spectrum of variations. For instance, to generate IM samples, one can simulate seismic wave propagation \cite{Rodgers2019BroadbandIntensities} to create synthetic seismograms or employ empirical ground motion prediction equations \cite{Boore2014NGA-West2Earthquakes}. Nonlinear structural dynamic analysis can be performed to generate EDP samples \cite{Zhang2023Regional-scaleIstanbul}, while several regression models directly connecting EDP to IM are also available \cite{Cornell2002ProbabilisticGuidelines,Heresi2021FragilityScale}. The DS can be obtained from structural analysis or inferred from IM or EDP samples using DS-IM or DS-EDP fragility curves. Lastly, DV can be assessed using loss curves that relate DS or EDP to DV \cite{Esmaili2018AnCounty}, or physics-based simulations \cite{Hulsey2022High-resolutionCordons}.

Physics-based simulations can accurately capture the complex interactions and correlations among structural responses and functionalities. However, they are often restricted to scenario-based analyses due to their extensive computational costs. On the other hand, statistical approaches, such as fragility function-based simulations of DS given EDP or IM \cite{Tabandeh2023SeismicFacilities}, are computationally feasible but require many ad-hoc, sometimes subjective, assumptions about statistical models. Specifically, applying structural-level fragility functions to regional-scale analysis entails constructing a \textit{fragility field} that summarizes the joint information across the region. To capture the spatial dependency among structural damage, the fragility field demands additional assumptions and parameters than individual fragility curves. The common practice of using individual fragility functions under the assumption of conditional independence (conditional on site-specific intensity measures) can significantly underestimate system-level failure probabilities \cite{Heresi2022Structure-to-structureAssessment}.

This study presents an alternative method for constructing the fragility field, which can be integrated with fragility-based approaches, physics-based simulations, and even post-hazard survey data. Specifically, we introduce the Ising model, a widely recognized model in statistical mechanics, as a joint probability mass function for binary damage states (safe or failure) of structures over a region. Beyond its applications in statistical mechanics, the Ising model has been used across various disciplines to examine collective behaviors in complex systems, including DNA sequence reconstruction, ecological relationships, and financial markets \cite{Nguyen2017InverseScience}. This versatility stems from its nature as the maximum entropy distribution for binary variables, subject to constraints on first- and second-order cross-moments. Recently, maximum entropy distributions have been applied to regional seismic simulation of road networks \cite{chu2023maximum}, where the focus is on the computational aspects of maximum entropy modeling and the use of an efficient surrogate model. This study positions the Ising model within the PBEE framework, concentrating on the meaningful interpretation of model parameters and simulation results. 

{\color{black}This paper is organized as follows}. Section \ref{Section:Current_approaches} overviews the current approaches for evaluating structural damage in the regional-scale PBEE. Subsequently, Section \ref{Section:Ising_model} introduces the formulation and parameter estimation of the Ising model in the context of regional seismic analysis. This section also develops the mean-field approximation to qualitatively explain the Ising model parameters. Section \ref{Section:Interpretation} offers interpretations of the Ising model parameters in the language of regional seismic risk analysis and explores their relationship with the mean and correlation coefficients of the damage states of structures. Section \ref{Section:NumericalExamples} demonstrates the validity and engineering relevance of the Ising model through two examples using real-world post-hazard survey data and synthetic simulation data, emphasizing the insights derived from the Ising model. Finally, concluding remarks and future research directions are presented in Section \ref{Section:Conclusions}.

\section{Current approaches for probabilistic structural damage predictions on a regional-scale} \label{Section:Current_approaches}
\noindent
For an individual structure, the following fragility function is commonly used to \textcolor{black}{estimate the probability of exceeding a damage state}:
    \begin{equation} \label{Eq:fragility}
    \Prob{DS\geq ds_i|IM=im}=\Phi \left(\frac{\ln\left(im/\theta_i\right)}{\beta_i}\right)\,,
    \end{equation}
where $ds_i$ represents the $i$-th damage state, $IM$ denotes the ground motion intensity at the location of the structure, $\Phi(\cdot)$ is the standard Gaussian cumulative distribution function, $\theta_i$ is the median of $IM$ causing the damage $ds_i$ or greater, and $\beta_i$ denotes the log standard deviation of $IM$ that induces the damage $ds_i$ or greater. In regional seismic simulations, it is a common practice to construct fragility functions for each structure and assume conditional independence between the damage states of structures. Upon generating spatially correlated ground motion intensity measures, damage states of the structures are sampled according to Eq.~\eqref{Eq:fragility} to produce a damage map \cite{Silva-Lopez2022DeepNetworks,Tabandeh2023SeismicFacilities,DeRisi2018Scenario-BasedScale,Silva-Lopez2022CommuterNetworks}. Subsequently, decision variables such as economic loss, casualties, increased traffic time, and welfare loss are calculated.
    
The missing information in the above procedure is the correlation between damage states contributed by the similarities of structures. \citet{Bazzurro2005AccountingEstimation} emphasized that accounting for the similarity of structures represents a key distinction between analyzing a single structure and a portfolio of structures in the seismic loss estimation. Similarities between structures, such as the number of stories, construction materials, and structural type, are particularly pronounced in residential neighborhoods where structures share similar designs and are built following the same codes. This contributes to the correlation between their damage states \cite{Heresi2022Structure-to-structureAssessment}, in addition to the contribution from the spatial coherency of ground motions. \textcolor{black}{Other than the damage states, the importance of accurately modeling the correlations between EDPs of structures was emphasized by \citet{KangEvaludationCorrleationEDP}, who proposed a framework to account for the correlations between EDP residuals, defined as the residual term in a regression function relating EDP to IM. This approach was later applied to regional seismic loss assessment \cite{Kang2022QuantifyingAssessment, Kang2023DeepStructures}. \citet{HamidiaPeakDrifRatio} and \citet{Asjodi3DFragilitySurface} also investigated the correlation between EDPs stemming from the similarities of mechanical properties, employing computer vision-based methods.}

To model the structure-to-structure damage correlation, \citet{Lee2007UncertaintySystems} presented a simplified approach by assuming equi-correlation among all structures. The study also concluded that the challenge of modeling the damage correlation lies in approximating the joint discrete damage distribution. Due to the difficulty of defining a joint distribution for discrete damage states, \citet{Baker2008UncertaintyEstimation} proposed using the joint continuous distribution for the decision variables given the engineering demand parameters, $f(\vect{dv}|\vect{edp})$, by integrating $f(\vect{dm}|\vect{edp})$ and $f(\vect{dv}|\vect{dm})$, thereby bypassing the need for a joint distribution of discrete damage states. An alternative approach, as detailed in \cite{Baker2008IntroducingComponents,Lee2021Multi-scaleAlgorithm}, models fragility using continuous demand and capacity. This method becomes popular because of its convenience.

The log-normal and other assumptions in fragility functions were largely motivated by convenience rather than by fundamental statistical and physical principles \cite{MasanobuShinozuka2000STATISTICALCURVES}. In the following sections, we will develop a joint distribution model for discrete damage states of structures that hinges on a single, fundamental assumption of statistical mechanics.

\section{Long-range Ising model as a joint probability mass function for damage states} \label{Section:Ising_model}
\noindent 
The Ising model was originally conceived as a mathematical model for ferromagnetism in statistical mechanics \cite{Kadanoff2000StatisticalRenormalization,Ma2019ModernPhenomena}. A key feature of the model is that it represents the maximum entropy distribution constrained by the means and pairwise correlations, making it applicable to a wide variety of fields beyond statistical mechanics \cite{Aurell2012InverseData, Schneidman2006WeakPopulation, Nguyen2017InverseScience}. Provided with the first- and second-order cross-moments of the component states in a system, the Ising model is the least structured joint probability distribution for the component states that reproduces the given information. In this section, we first formulate the Ising model within the context of regional seismic analysis, followed by a discussion on its parameter estimation. Finally, we present the mean-field approximation, which offers a qualitative understanding of the relationship between the Ising model parameters and the given statistics.
\subsection{Formulation} \label{Formulation}
\noindent 
Consider a region consisting of $n$ structures under an earthquake and assume that each structure can be in one of the binary states of safe or failure. Here, safe and failure are relative terms with respect to a target performance level. The group of structures has $N=2^n$ configurations, where the probability of the $I$-th configuration is expressed as:
    \begin{equation}
    P_I = \Prob{X_1=x_{1I}, X_2=x_{2I},\dots, X_n = x_{nI}}\,,    
    \end{equation}    
where $X_i\in\{-1\text{ (safe)},1\text{ (failure)}\}$ is a random variable representing the state of the $i$-th structure, and $x_{iI}$ denotes an outcome of $X_i$ for the $I$-th configuration. 

The entropy $S$ for $\vect X=[X_1,X_2,...,X_n]\tr$ is defined as:
    \begin{equation} \label{Eq:entropy}
    S\defi-\sum_{I=1}^N P_I \log{P_I}.
    \end{equation}
In the absence of information, the probability distribution that maximizes the entropy is the uniform distribution, which means that all possible configurations are equally likely. Suppose we have information up to the second-order cross-moments expressed as follows:
    \begin{align}
    &\mathcal{M}_{1,i} = \sum_{I=1}^N x_{iI}P_I\,, \label{Eq:1storder}\\
    &\mathcal{M}_{2,ij} = \sum_{I=1}^N x_{iI}x_{jI}P_I\,,i,j=1,2,...,n\,, \label{Eq:2ndorder}
    \end{align}
where $\mathcal{M}_{1,i}$ is the first-order cross-moment, i.e., the mean of $X_i$, and $\mathcal{M}_{2,ij}$ is the second-order cross-moment between the damage states of the $i$-th and $j$-th structures. Applying Lagrange multipliers with $\mathcal{M}_{1,i}$ and $\mathcal{M}_{2,ij}$ as constraints, the joint probability $P_I$ that maximizes the entropy $S$ is:
    \begin{equation} \label{Eq:Isingmodel_temp}
    P_I=\frac{1}{Z}\exp{\left(\sum_{i=1}^{n}h_ix_{iI}+\sum_{i>j}J_{ij}x_{iI}x_{jI}\right)}\,,
    \end{equation}
where $Z$ is the normalizing constant given as: 
    \begin{equation} \label{Eq:PartitionFunction}        Z=\sum_{I=1}^{N}\exp{\left(\sum_{i=1}^{n}h_ix_{iI}+\sum_{i>j}J_{ij}x_{iI}x_{jI}\right)}\,,
    \end{equation}
where $h_i$ and $J_{ij}$ are the Lagrange multipliers to meet the first- and second-order cross-moment constraints, respectively. The detailed derivation is presented in \ref{Appendix:Derivation}. Since our goal is to model the joint distribution that outputs  the probability for any specified configuration, we drop the subscript $I$ in Eq.~\eqref{Eq:Isingmodel_temp} and replace $P_I$ by $p(\vect x)$:
\begin{equation}
   \begin{aligned} \label{Eq:Isingmodel}
    p(\vect x) &= \frac{1}{Z}\exp{\left(\sum_{i=1}^{n}h_ix_{i}+\sum_{i>j}J_{ij}x_{i}x_{j}\right)} \\
      &= \frac{1}{Z}\exp{\left(\vect{h}\tr\vect{x}+\frac{1}{2}\vect{x}\tr\vect{J}\vect{x}\right)}\,,
    \end{aligned} 
\end{equation}
 where $\vect{x}=\left[x_1,x_2,\dots,x_n\right]\tr$, $\vect{h}=\left[h_1,h_2,\dots,h_n\right]\tr$, and: 
    \begin{equation}
   \vect{J}=\begin{bmatrix}
    0      & J_{12} & \dots  & J_{1n} \\
    J_{12} & 0      & \dots  & J_{2n} \\
    \vdots & \vdots & \ddots & \vdots \\
    J_{1n} & J_{2n} & \dots  & 0
    \end{bmatrix}\,.      
    \end{equation}
Notice that Eq.~\eqref{Eq:Isingmodel} and Eq.~\eqref{Eq:Isingmodel_temp} are equivalent, because any configuration of $\vect x$ is indexed by $I$. Eq.~\eqref{Eq:Isingmodel} is the Ising model for describing ferromagnetism, where $h_i$ and $J_{ij}$ can be interpreted as the external magnetic field acting on each component/spin and the pairwise coupling effects between each spin pair, respectively \cite{Kadanoff2000StatisticalRenormalization,Ma2019ModernPhenomena}. The classic Ising model features a sparse $\vect J$ because it considers only the pairwise coupling effects between neighboring spins on a lattice. However, in this study,  Eq.~\eqref{Eq:Isingmodel} represents a \textit{long-range} or \textit{fully-connected} Ising model, accounting for interactions between all pairs of structures in a region.

\subsection{Parameter estimation} \label{Subsection:ParameterEstimation}
\noindent
Parameter estimation of the Ising model, known as the \textit{inverse Ising problem}, is challenging for large systems comprising numerous components. Existing computational techniques have yet to fully overcome the curse of dimensionality in solving inverse Ising problems involving over $10{,}000$ components \cite{broderick2007faster,Sohl-Dickstein2011NewFlow,Obermayer2014InverseSamples,Cocco2018InverseReview}. Nonetheless, for systems with hundreds of components, a variety of convex optimization algorithms are effective, owing to the convexity of the inverse Ising problem \cite{Nguyen2017InverseScience}.

The Boltzmann machine learning, a widely used method for the inverse Ising problem, obtains the maximum likelihood estimation of the Ising model parameters through gradient-based optimization. The log-likelihood $\mathcal{L}$ of the parameters $\vect{h}$ and $\vect{J}$ given \textcolor{black}{a collection of $M$ observations, $\mathcal{D}=\{\vect{x}^{(1)},\vect{x}^{(2)},\ldots,\vect{x}^{(M)}\}$,} is expressed as:
\begin{equation}
\begin{aligned}
 \mathcal{L}(\vect{h},\vect{J}|\mathcal{D}) & = \frac{1}{M}\ln{\left(\prod_{m=1}^{M} p(\vect{x}^{(m)}|\vect{h},\vect{J}) \right)} \\
                                & = \sum_{i=1}^n h_i\left({\color{black}{\frac{1}{M}\sum_{m=1}^M x^{(m)}_i}} \right)+\sum_{i>j} J_{ij}\left({\color{black}{\frac{1}{M}\sum_{m=1}^M x^{(m)}_i x^{(m)}_j}} \right)-\ln{Z(\vect{h},\vect{J})} \\
                                & = \sum_{i=1}^n h_i \mathcal{M}_{1,i}^\mathcal{D}+\sum_{i>j} J_{ij}\mathcal{M}_{2,ij}^\mathcal{D}-\ln{Z(\vect{h},\vect{J})} \,,\label{Eq:LogLikelihood}          
\end{aligned}
\end{equation}
where $\mathcal{M}_{1,i}^\mathcal{D}$ and $\mathcal{M}_{2,ij}^\mathcal{D}$ denote the first- and second-order cross-moments calculated from the observations $\mathcal{D}$. The final line of Eq.~\eqref{Eq:LogLikelihood} reveals that the likelihood function hinges on $\mathcal{M}_{1,i}^\mathcal{D}$ and $\mathcal{M}_{2,ij}^\mathcal{D}$ rather than on the random samples in $\mathcal{D}$. This suggests that the Ising model can be applied when $\mathcal{M}_{1,i}^\mathcal{D}$ and $\mathcal{M}_{2,ij}^\mathcal{D}$ are derived from sources other than random samples. Consequently, the Ising model can also be constructed using fragility functions developed either from empirical data \cite{Straub2008ImprovedData,Rosti2021EmpiricalBuildings,DelGaudio2017EmpiricalEarthquake}, analytical models \cite{avcsar2011analytical,Rota2010,Ji2007AnBuildings}, or a combination of both \cite{Anagnos1995NCEER-ATCBuildings}, without requiring information beyond the current PBEE practice.

Notice that $\ln{Z}$ serves as a generating function for the first- and second-order cross-moments: $\frac{\partial \ln{Z}}{\partial h_i}=\mathcal{M}_{1,i}$ and $\frac{\partial \ln{Z}}{\partial J_{ij}}=\mathcal{M}_{2,ij}$. Using this property, differentiating Eq.~\eqref{Eq:LogLikelihood} with respect to $h_i$ and $J_{ij}$ yields:
\begin{equation}
     \begin{aligned}
    &\frac{\partial \mathcal{L}}{\partial h_i}(\vect{h},\vect{J}) = \mathcal{M}_{1,i}^\mathcal{D}-\mathcal{M}_{1,i}\,,\\
    &\frac{\partial \mathcal{L}}{\partial J_{ij}}(\vect{h},\vect{J}) = \mathcal{M}_{2,ij}^\mathcal{D}-\mathcal{M}_{2,ij}\,.\label{Eq:gradientdescent}
    \end{aligned}
\end{equation}
Finally, we obtain \textcolor{black}{the gradient descent updating rule for the $k$-th step of} the maximum likelihood estimation:
\begin{equation}
    \begin{aligned} 
    &h_i^{(k+1)} = h_i^{(k)}+\alpha \frac{\partial \mathcal{L}}{\partial h_i}(\vect{h}^{(k)},\vect{J}^{(k)})=h_i^{(k)}+\alpha\left(\mathcal{M}_{1,i}^\mathcal{D}-\mathcal{M}_{1,i}\right)\,,\\
    &J_{ij}^{(k+1)} = J_{ij}^{(k)}+\alpha \frac{\partial \mathcal{L}}{\partial J_{ij}}(\vect{h}^{(k)},\vect{J}^{(k)})=J_{ij}^{(k)}+\alpha\left(\mathcal{M}_{2,ij}^\mathcal{D}-\mathcal{M}_{2,ij}\right)\,,\label{Eq:updating}
    \end{aligned}
\end{equation}
where $\alpha$ is the learning rate of the algorithm. \textcolor{black}{The initial values are set to $h_i^{(1)}=\tanh^{-1}\mathcal{M}_{1,i}$ and $J_{ij}^{(1)}=0$, based on the mean-field solution with zero pairwise interaction, which will be discussed in the following subsection}. Owing to the convexity of the inverse Ising problem, Eq.~\eqref{Eq:updating} is guaranteed to converge to a unique solution, \textcolor{black}{i.e., a unique set of Ising model parameters that reproduces the provided information} \cite{Nguyen2017InverseScience}. However, the exact evaluation of $\mathcal{M}_{1,i}$ and $\mathcal{M}_{2,ij}$ at each step requires calculating Eqs.~\eqref{Eq:1storder} and \eqref{Eq:2ndorder} over $2^n$ configurations, which is generally infeasible. Hence, Monte Carlo sampling techniques are often used to approximate $\mathcal{M}_{1,i}$ and $\mathcal{M}_{2,ij}$ \cite{Habeck2014BayesianMechanics}. In this study, we adopted Gibbs sampling due to its applicability to high-dimensional distributions and ease of implementation.

\subsection{Mean-field approximation} \label{Subsection:mean-field approximation}
\noindent
In this subsection, we introduce the mean-field approximation, one of the simplest approximation methods \cite{Kappen1998EfficientTheory, Roudi2009StatisticalModels, Nguyen2012Mean-fieldTemperatures, Nguyen2017InverseScience} for the inverse Ising problem, to facilitate the interpretation of the Ising model parameters. By averaging the interactions between components, the mean value of each component for a specified $\vect{h}$ and $\vect{J}$ is given by the solution of the following \textit{self-consistent equation}:
    \begin{equation} \label{Eq:meanfield_m_forward}
    m_i = \tanh\left({h_i+\sum_{j\neq i}J_{ij}m_{j}}\right)\,,
    \end{equation}
in which $m_i=\mathcal{M}_{1,i}$ denotes the mean value of the $i$-th component. The covariance can be obtained by solving:
    \begin{equation} \label{Eq:meanfield_c_forward}
    c_{ij} = \left(1-m_i^2\right)\left(\delta_{ij}+\sum_{k\neq i}J_{{\color{black}{ik}}}c_{kj}\right)\,,
    \end{equation}
where $c_{ij}$ is the covariance between the $i$-th and $j$-th components, which is equivalent to $\mathcal{M}_{2,ij}-\mathcal{M}_{1,i}\mathcal{M}_{1,j}$, and $\delta_{ij}$ is the Kronecker delta. The detailed derivation for the mean-field approximation is provided in \ref{Appendix:Meanfieldsolution}. Eqs.~\eqref{Eq:meanfield_m_forward} and \eqref{Eq:meanfield_c_forward} represent the \textit{forward} approximation that derives $m_i$ and $c_{ij}$ from the given $\vect h$ and $\vect J$. For the inverse Ising problem, the mean-field solution derived from the forward approximation is:
\begin{equation}
     \begin{aligned}
    &h_i    = \tanh^{-1}{m_i}-\sum_{j\neq i}J_{ij}m_{j}\,,\\
    &J_{ij} = -\left(\vect{c}^{-1}\right)_{ij}\,, 
    \end{aligned}\label{Eq:meanfield_inverse}
\end{equation}
where $\vect{c}$ is the covariance matrix with $c_{ij}$ as elements.

The accuracy and validity of the mean-field approximation depend on the magnitude and homogeneity of the pairwise interactions, as this method averages these interactions and neglects individual effects. \textcolor{black}{Notably, the mean-field solution becomes exact when the damage states are independent, i.e., for all pairs, $J_{ij}=0$ }. Although various methods have been proposed to improve the applicability of the mean-field approximation, such as the Thouless-Anderson-Palmer (TAP) method \cite{Kappen1998EfficientTheory, Nguyen2017InverseScience} and the Sessak-Monasson expansion \cite{Sessak2009Small-correlationProblem}, we found that none of these methods achieve satisfactory accuracy in regional seismic analysis \textcolor{black}{with non-negligible correlations}. Therefore, our usage of the mean-field approximation is to facilitate a qualitative interpretation of the model parameters. 

\section{Parametric study of the Ising model} \label{Section:Interpretation}
\noindent 
Using simple proof-of-concept examples, this section investigates the relationships between the Ising model parameters and typical performance quantities, such as failure probability and the correlation coefficient between damage states of structures.
\subsection{Interpretations of the Ising model parameters}
\noindent 
\begin{itemize}
\item $\vect h$: We interpret $\vect h$ as a \textit{risk field} acting on each structure within a region. According to Eq.~\eqref{Eq:Isingmodel}, a structure with a positive $h_i$ {\color{black} favors the state of failure}, i.e., to have a spin of $x_i=+1$, with a larger $h_i$ signifying a greater tendency toward failure. Conversely, a negative risk field $h_i$ indicates a tendency to remain safe. Given that civil structures are engineered to \textcolor{black}{remain reliable in the absence of hazards, they should inherently possess a negative field $-\mathcal{I}$, where $\mathcal{I}>0$ can be interpreted as the inertia toward a safe state}. A hazard imposes an external positive field $h_H>0$ on each structure, and the resulting risk field is a balance between the hazard and the inertia, i.e., $h_i=h_H-\mathcal{I}$\textcolor{black}{, where both $h_H$ and $\mathcal{I}$ vary across structures. Here, the decomposition of $h_i$ into $h_H$ and $\mathcal{I}$ is introduced to facilitate a deeper understanding, not for any computational purpose}.
\item $\vect J$: This represents the \textit{pairwise interaction}, reflecting the tendency of two structures to be in the same damage state. The term \textit{interaction} encompasses all direct and indirect effects that determine how the damage state of one structure can affect the state of another\textcolor{black}{, generated by statistical correlations and (potentially) physical contact actions}. From Eq.~\eqref{Eq:Isingmodel}, {\color{black}it is observed that a positive $J_{ij}$ favors a synchronization between states, i.e., the corresponding pair of structures have states $(+1,+1)$ or $(-1,-1)$, with a larger $J_{ij}$ increasing this tendency}. On the other hand, a negative $J_{ij}$ favors the damage states being different. 
\end{itemize} 

The specific choice of terminology, such as \textit{risk field} and \textit{pairwise interaction}, can vary, but the fundamental implications of $\vect h$ and $\vect J$ remain invariant. Recall from Section \ref{Section:Ising_model} that the construction of the Ising model relies solely on the first- and second-order cross-moments, \textcolor{black}{which might mislead one to argue that the Ising model cannot offer anything beyond the mean and the covariance}. However, it is important to acknowledge that the principle of maximum entropy introduces a crucial additional layer of information. Therefore, the Ising model encompasses more than just the mean and the covariance matrix\textcolor{black}{, allowing inferences about the regional seismic behavior that is not apparent from the data.}

\subsection{Dependency between the Ising model parameters and the mean and correlation coefficient} \label{Subsection:Correspondence_to_mean_and_correlation}
\noindent 
Consider two binary random variables $\vect X=[X_1,X_2]\tr\in\{-1\text{(safe)},1\text{(failure)}\}^2$ representing the damage states of two buildings. The risk field and pairwise interaction are given as $h_1=h_2=h$ and $J_{12}=J$, respectively. With only 2 variables and thus $2^2$ system configurations, the exact mean and correlation coefficient for given $h$ and $J$ values can be \textcolor{black}{computed} using Eqs.~\eqref{Eq:1storder}-\eqref{Eq:Isingmodel}. The mean value $\E{X_i}$ is directly linked to the failure probability $P_f$ by $P_f=(\E{X_i}+1)/2$. Note that, due to parameter settings of the current demonstration, the failure probability $P_f$ for each building is the same. 

\begin{figure}[t]
\begin{subfigure}[T]{.49\textwidth}
    \centering
    \includegraphics[width=1.0\linewidth]{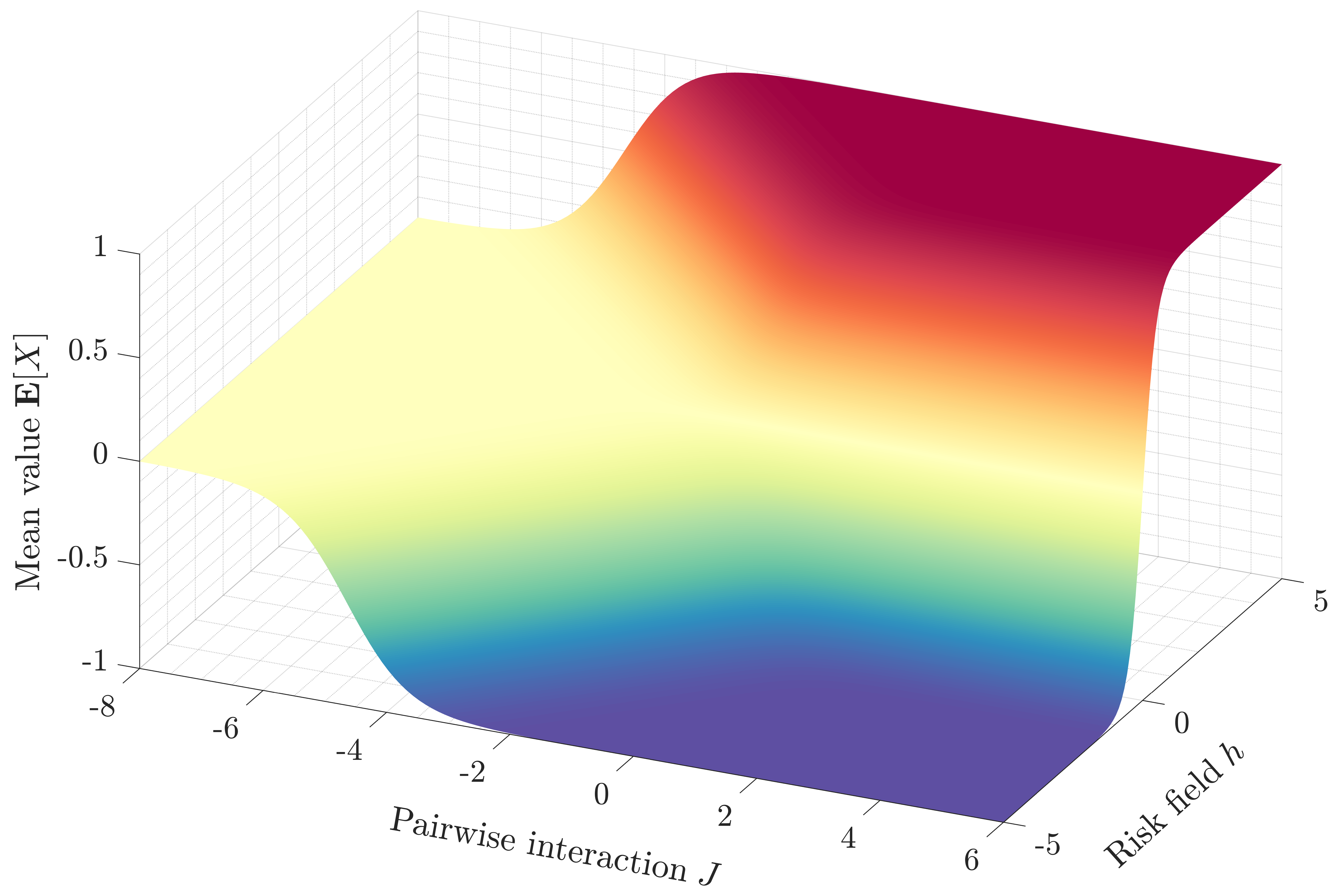}
    \caption{\textbf{}}
    \label{Fig:Correspondence_mean_3d}
\end{subfigure}%
\begin{subfigure}[T]{.49\textwidth}
    \centering
    \includegraphics[width=1\linewidth]{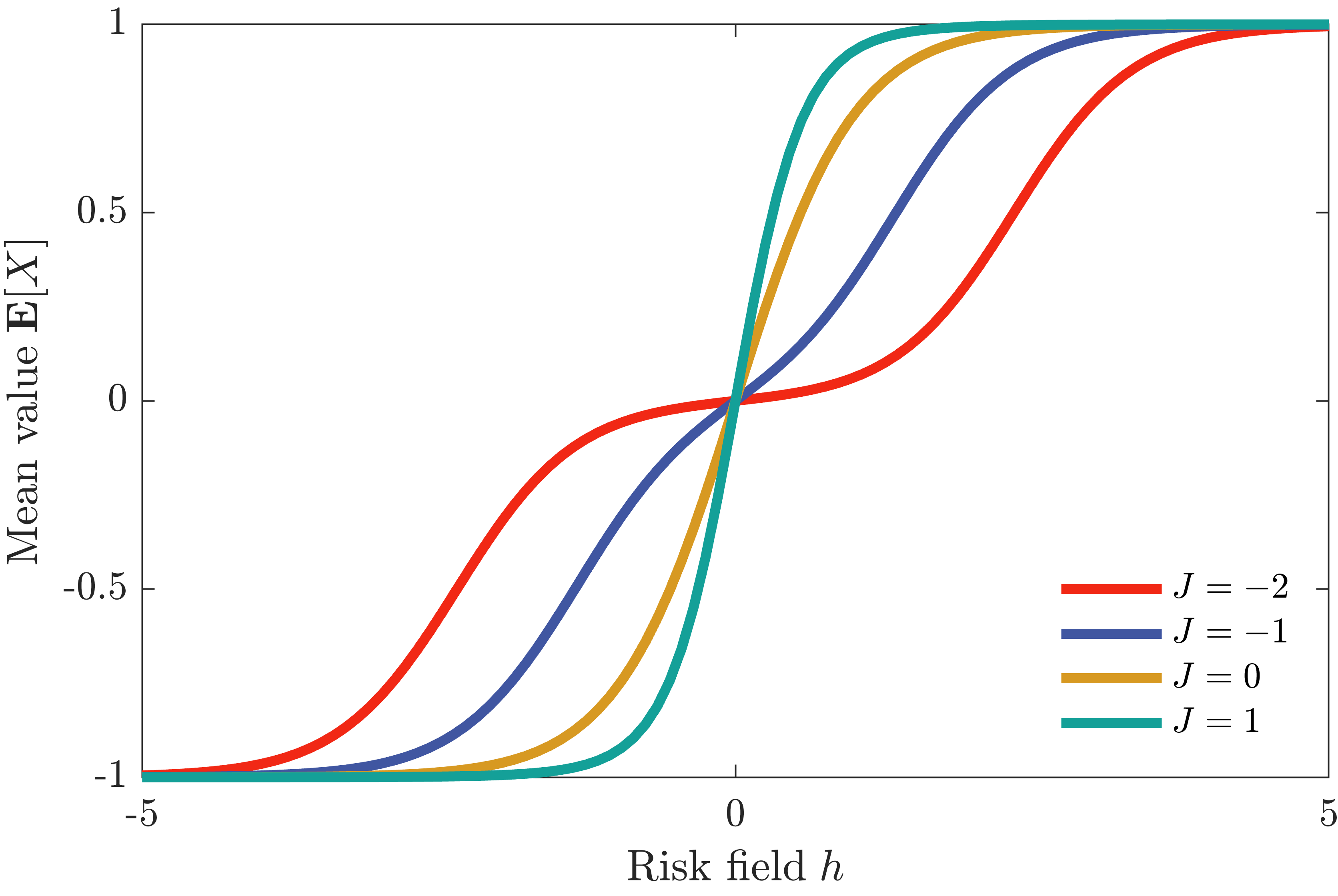}
   \caption{\textbf{}}
    \label{Fig:Correspondence_mean_h}
\end{subfigure}
\caption{\textbf{The relationship between the mean $\E{X_i}$, risk field $h$, and pairwise interaction $J$ for two buildings: (a) three-dimensional representation, and (b) two-dimensional illustration of $\E{X_i}$ against $h$ for four $J$ values}. For any $J$, the mean damage state $\E{X_i}$ is monotonically increasing with respect to $h$. A negative interaction $J$ flattens the increase, while a positive $J$ steepens it.}
\label{Fig:Correspondence_mean}
\end{figure}

The mean value for the damage state of a building is affected both by the risk field acting on the site and the pairwise interaction from the other building. Figure \ref{Fig:Correspondence_mean_3d} depicts the mean damage state of a building with varying $h$ and $J$, and Figure \ref{Fig:Correspondence_mean_h} illustrates the relationship between the mean value and the risk field $h$ under four different $J$ values. The mean damage state of a building becomes larger as the risk field $h$ increases, reflecting that a stronger risk field results in a higher failure probability. {\color{black} Importantly, Figure \ref{Fig:Correspondence_mean_h} also suggests that: (i) when $h>0$, negative pairwise interaction provides a buffer against failure, i.e., to achieve the same mean damage $\E{X_i}>0$, negative $J$ requires larger $h$; while (ii) when $h<0$, positive pairwise interaction provides a buffer against failure, i.e., to achieve the same $\E{X_i}<0$, positive $J$ requires larger $h$. This observation has a clear interpretation: (i) when the hazard is weak, adding positive correlation into the system increases the likelihood of a \textit{collective safe state} that helps to resist failure; while (ii) when the hazard is strong, adding positive correlation triggers a \textit{collective failure state} that is detrimental. In short, coherency can be both a shield and a vulnerability. Incidentally, the change of behaviors with flipping the sign of $h$ is associated with the \textit{first-order phase transition} in statistical mechanics. We will examine first- and second-order phase transitions within regional seismic responses in upcoming research.}

\begin{figure}[H]
\begin{subfigure}[t]{.49\textwidth}
    \centering
    \includegraphics[width=1.0\linewidth]{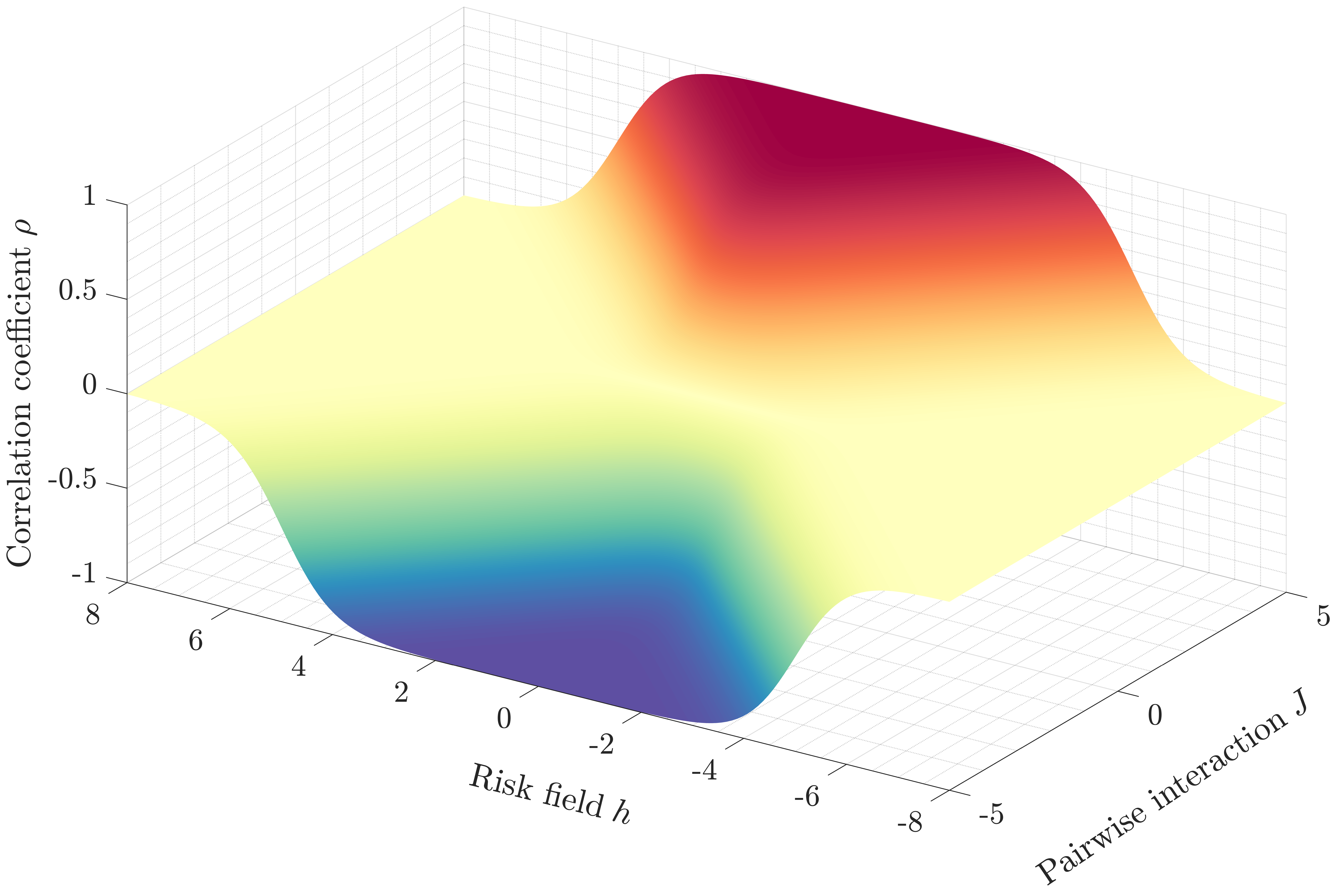}
    \caption{\textbf{}}
    \label{Fig:Correspondence_correlation_3d}
\end{subfigure}%
\begin{subfigure}[t]{.49\textwidth}
    \centering
    \includegraphics[width=1.0\linewidth]{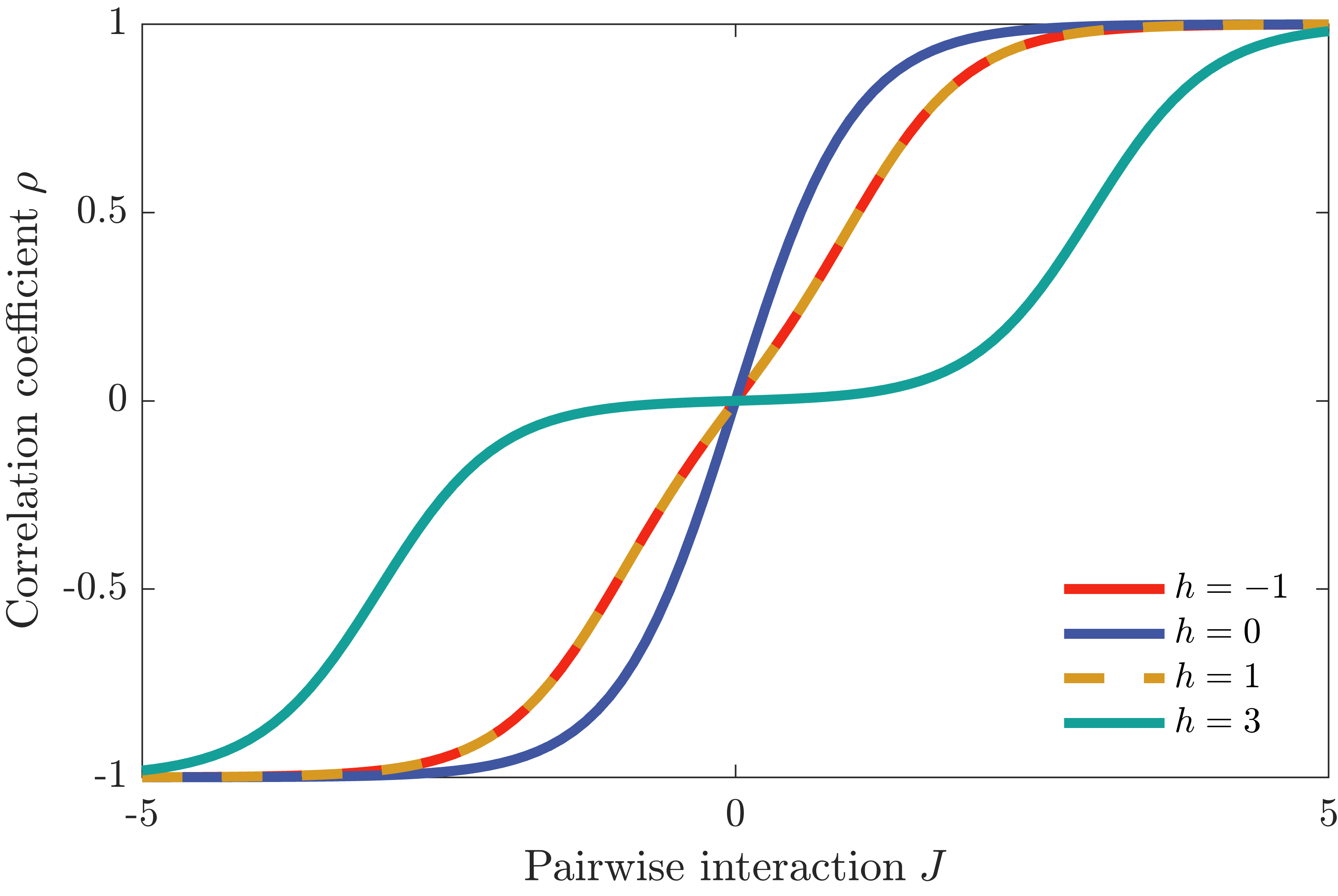}
    \caption{\textbf{}}
    \label{Fig:Correspondence_correlation_J}
\end{subfigure}
\caption{\textbf{The relationship between the correlation coefficient $\rho$, risk field $h$, and pairwise interaction $J$ for two buildings: (a) three-dimensional representation, and (b) two-dimensional illustration of $\rho$ against $J$ for four $h$ values}. For any $h$, the correlation coefficient $\rho$ is monotonically increasing with respect to $J$. A large $|h|$ flattens the increase, while a small $|h|$ steepens it.}
\label{Fig:Correspondence_correlation}
\end{figure}

Figures \ref{Fig:Correspondence_correlation} reveals that the correlation coefficient of the damage states is influenced by both the pairwise interaction $J$ and the risk field $h$. For a given $h$, the correlation coefficient is monotonically increasing with respect to $J$. Increasing the absolute value of $h$ will make the $\rho$-$J$ dependency weaker. This occurs because, for large $|h|$, the joint distribution is dominated by the term $\sum_i h_ix_i$, rather than $\sum_{i>j}J_{ij}x_ix_j$, and the term $\sum_i h_ix_i$ implies independence. Moreover, the sign of $h$ will not influence the $\rho$-$J$ dependency for this simple model with two structures; this particular observation cannot be generalized to long-range Ising models with nonhomogeneous parameters. 

\begin{figure}[H]
\begin{subfigure}[t]{.49\textwidth}
    \centering
    \includegraphics[width=0.5\linewidth]{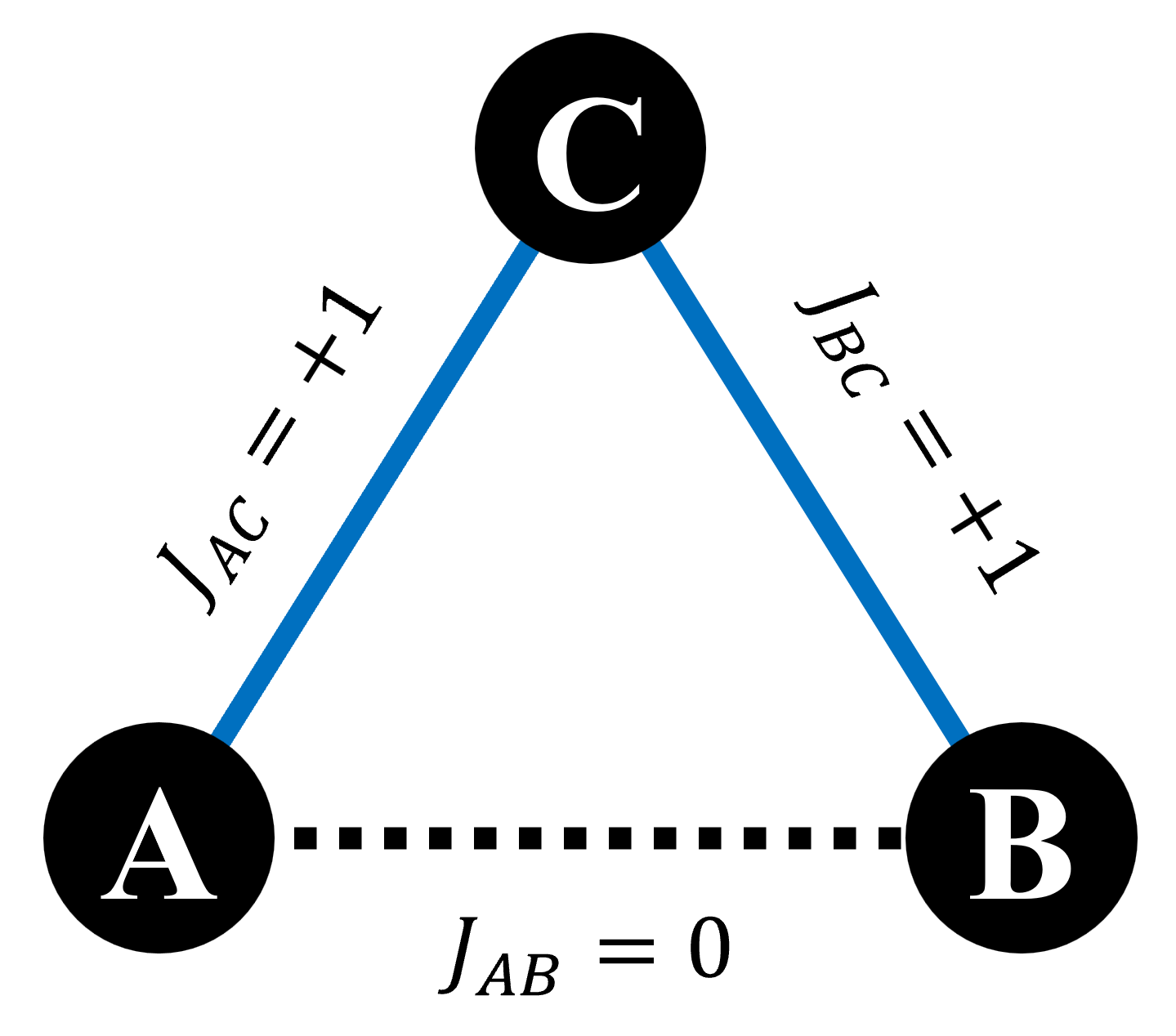}
    \caption{\textbf{}}
    \label{Fig:ThreeNodes_1_J}
\end{subfigure}%
\begin{subfigure}[t]{.49\textwidth}
    \centering
    \includegraphics[width=0.5\linewidth]{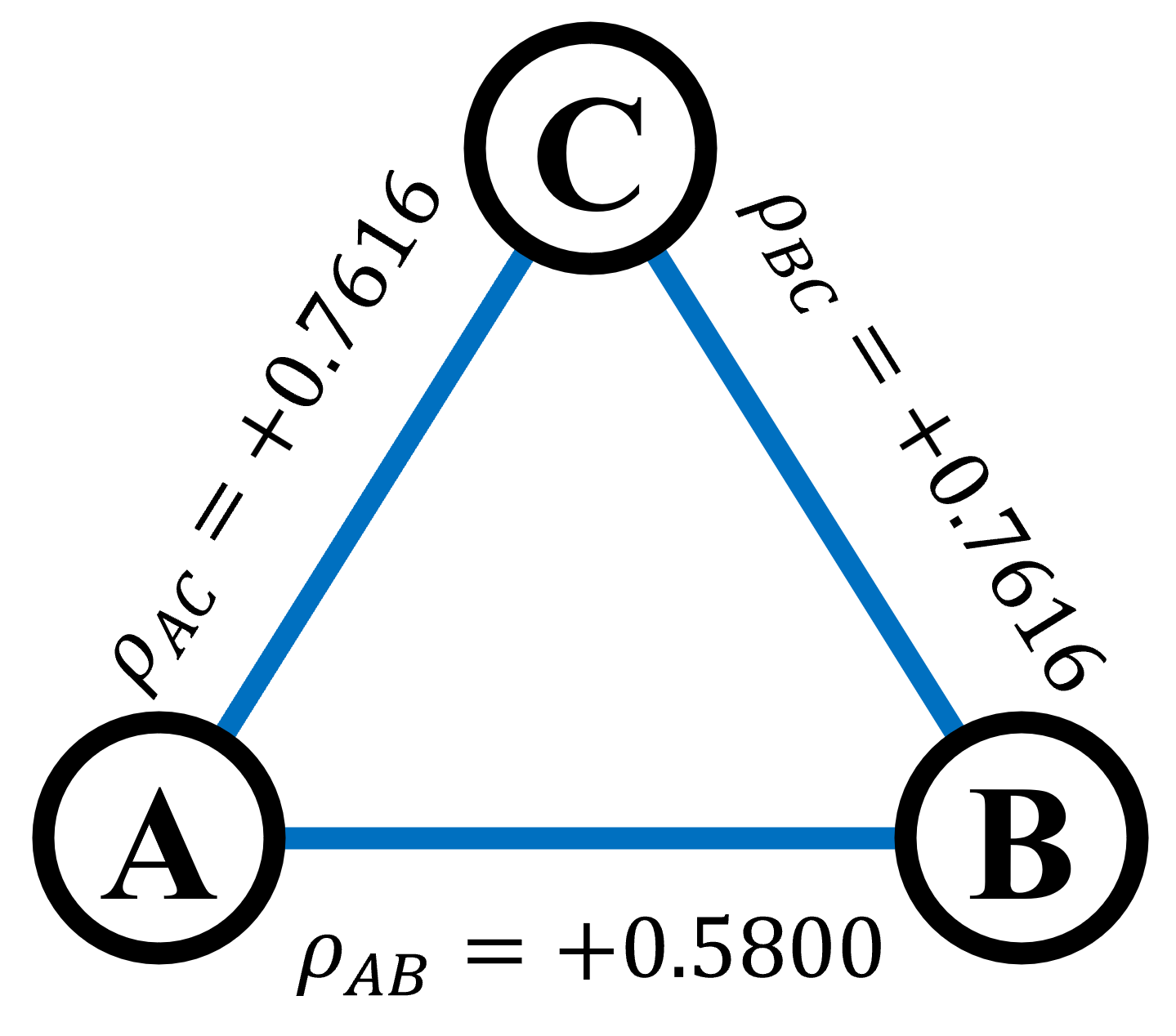}
    \caption{\textbf{}}
    \label{Fig:ThreeNodes_1_rho}
\end{subfigure}
\caption{\textbf{A 3-node system with zero pairwise interaction between nodes A and B: (a) the pairwise interaction values, and (b) the corresponding correlation coefficient values}. A non-zero correlation coefficient does not necessarily imply a non-zero pairwise interaction.}
\label{Fig:ThreeNodes_1}
\end{figure}

\begin{figure}[H]
\begin{subfigure}[t]{.49\textwidth}
    \centering
    \includegraphics[width=0.5\linewidth]{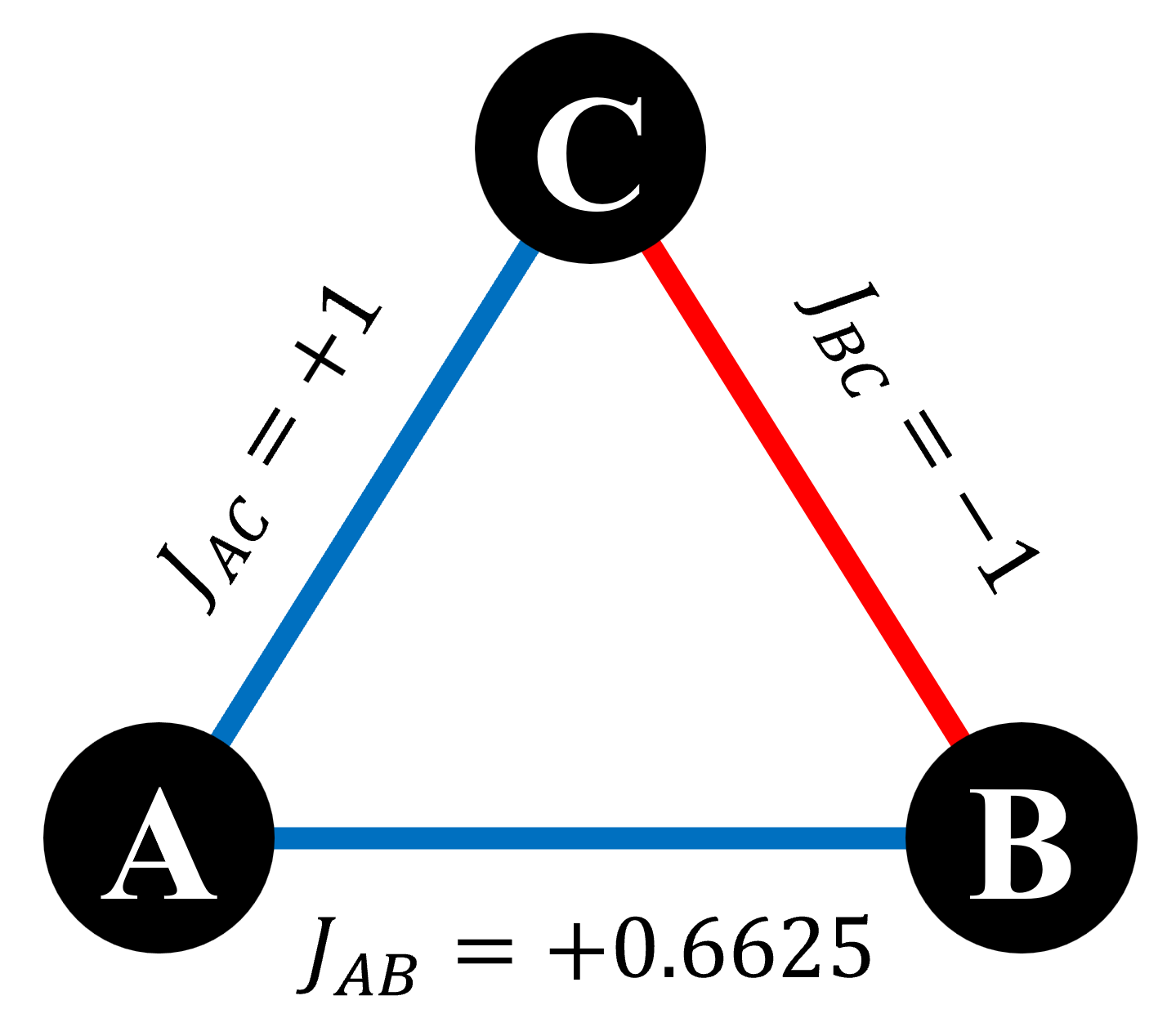}
    \caption{\textbf{}}
    \label{Fig:ThreeNodes_2_J}
\end{subfigure}%
\begin{subfigure}[t]{.49\textwidth}
    \centering
    \includegraphics[width=0.5\linewidth]{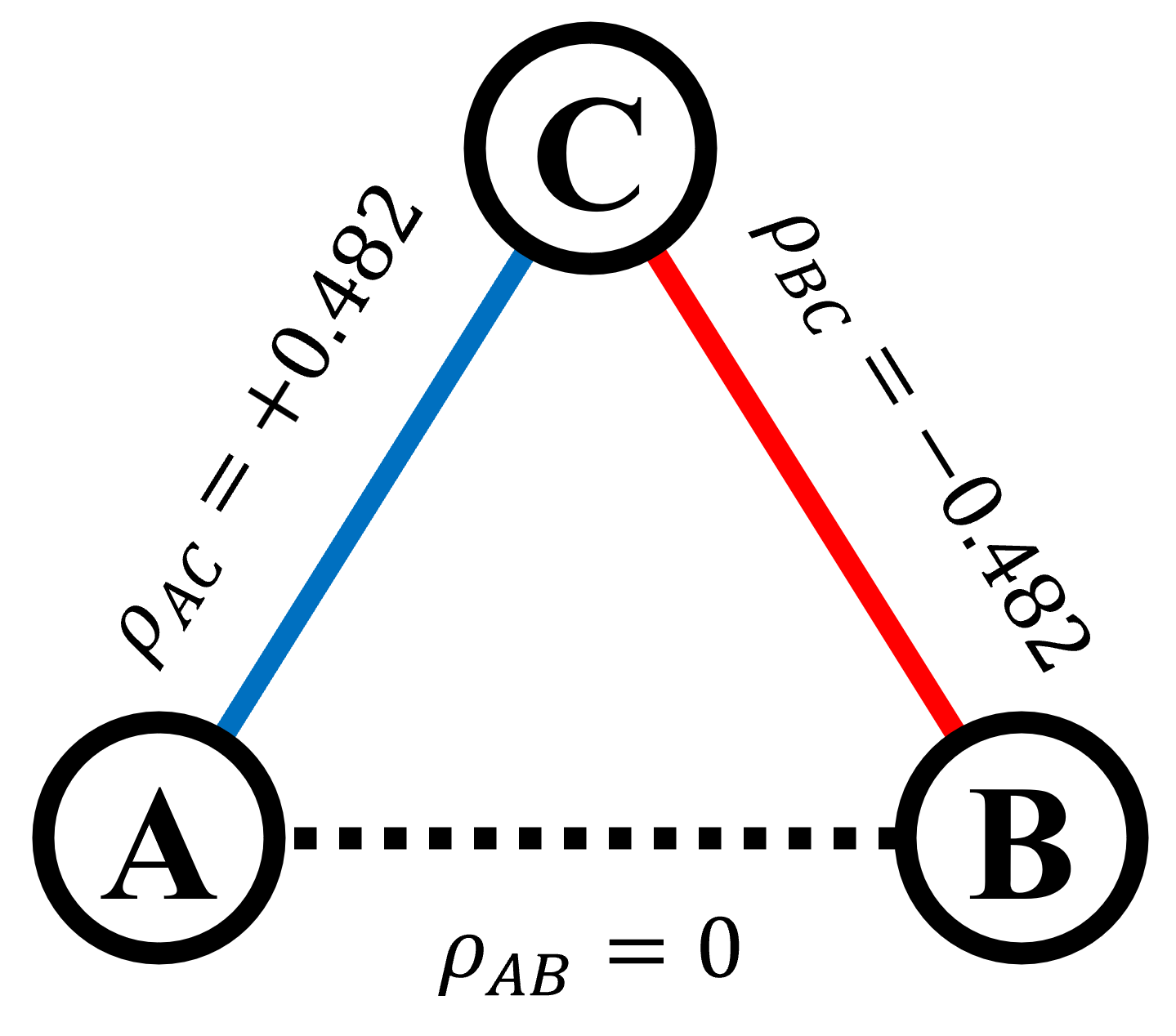}
    \caption{\textbf{}}
    \label{Fig:ThreeNodes_2_rho}
\end{subfigure}
\caption{\textbf{A 3-node system with zero correlation coefficient between nodes A and B: (a) the pairwise interaction values, and (b) the corresponding correlation coefficient values}. A zero correlation coefficient does not necessarily suggest a zero pairwise interaction.}
\label{Fig:ThreeNodes_2}
\end{figure}

To further investigate the relationship between the correlation coefficient and pairwise interaction, we consider a scenario involving three buildings without a risk field. Figure \ref{Fig:ThreeNodes_1} shows the $J_{ij}$ values for each pair of buildings and the corresponding correlation coefficient values. Despite having no direct interaction, i.e., $J_{AB}=0$, buildings A and B exhibit a correlation coefficient of $0.58$. This is attributed to the significant pairwise interactions between buildings A and C and between buildings C and B. Additionally, Figures \ref{Fig:ThreeNodes_2} demonstrates that buildings A and B have a zero correlation coefficient, despite a large positive $J$ value of $0.66$. This occurs because the indirect effects of the positive interaction between buildings A and C and the negative interaction between buildings B and C outweigh the direct interaction between buildings A and B.        

The illustrations above demonstrate the complex interplay between the Ising model parameters and both the mean and the correlation coefficient. The mean damage state of a structure is primarily influenced by the risk field acting on it, although the pairwise interactions also contribute to either sharpening or diminishing this effect. Similarly, the correlation coefficient for the damage states between a pair of structures is influenced by all pairwise interactions involving those structures, with the magnitude of the risk field either amplifying or mitigating this effect. These dynamics are approximately represented by the mean-field solutions in Eqs.~\eqref{Eq:meanfield_m_forward} and \eqref{Eq:meanfield_c_forward}. In essence, the Ising model, parameterized by $\vect{h}$ and $\vect{J}$, provides a fundamental mechanism for generating the mean damage states and correlation coefficients.

\section{Numerical examples} \label{Section:NumericalExamples}
\noindent 
 This section investigates the engineering relevance of the Ising model in the context of regional seismic analysis. The data for constructing the Ising model, the first- and second-order cross-moments for damage states of structures, can be obtained from fragility curves and their correlation models \cite{Baker2008IntroducingComponents}. This data can also be generated from regional seismic response simulators, such as the Regional Resilience Determination (R2D) tool of SimCenter \cite{McKenna2024NHERI-SimCenter/R2DTool:4.0.0} and the Interdependent Networked Community Resilience Modeling Environment (IN-CORE) platform of NCSA \cite{vandeLindt2023TheIN-CORE}. In this study, we first demonstrate how the Ising model can be built using post-earthquake damage evaluation data from a region in Antakya, Turkey, following the 2023 Turkey-Syria earthquake. Next, we construct the Ising model for a neighborhood in Pacific Heights, San Francisco, using samples generated by the R2D tool. 

\subsection{Example 1: Ising model from post-earthquake damage evaluation data}
\begin{figure}[H]
    \centering
    \includegraphics[height=6.5cm,trim={0 0.7cm 0 1cm},clip]{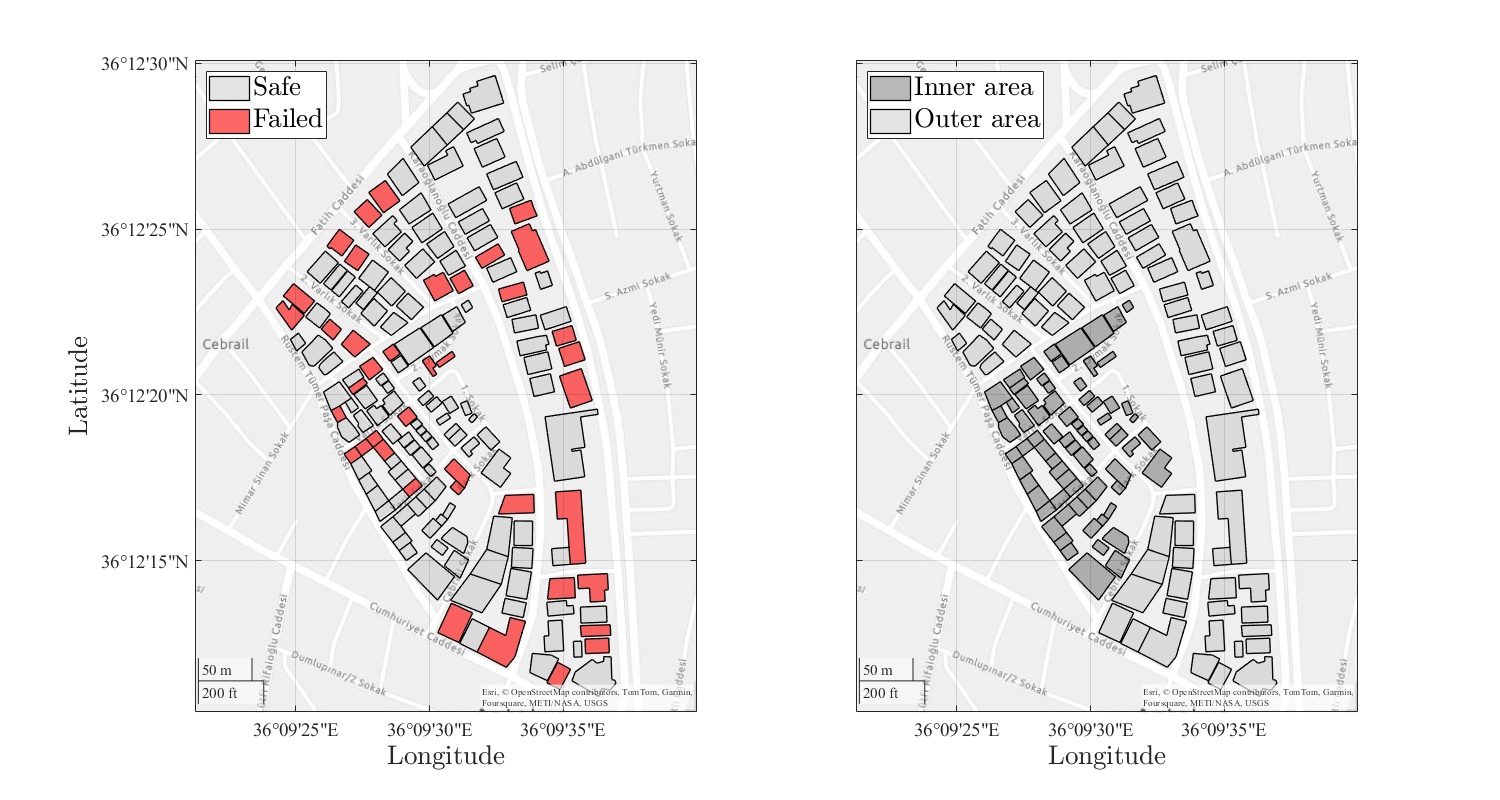}
    \caption{\textbf{The building damage in a region of Antakya, Turkey, following the 2023 Turkey-Syria earthquake}. \textcolor{black}{Out of 156 buildings, 40 failed, including 14 out of 72 failing in the inner area and 26 out of 84 failing in the outer area. The inner and outer areas are defined in the right panel; this definition will facilitate the verification of the Ising model predictions}. }
    \label{Fig:Turkey_observation}
\end{figure}

\noindent
A neighborhood in Antakya, Turkey, is selected as the target region for constructing the Ising model. The data on building damage is obtained from the Humanitarian OpenStreetMap (HOTOSM) Team \cite{Turkey_buildings,Turkey_destroyed_buildings}. Figure \ref{Fig:Turkey_observation} illustrates the damage states for the target region following the 2023 Turkey-Syria earthquake. The region consists of 156 buildings, 40 of which failed during the earthquake.

\begin{figure}[H]
    \centering
    \includegraphics[width=0.5\linewidth]{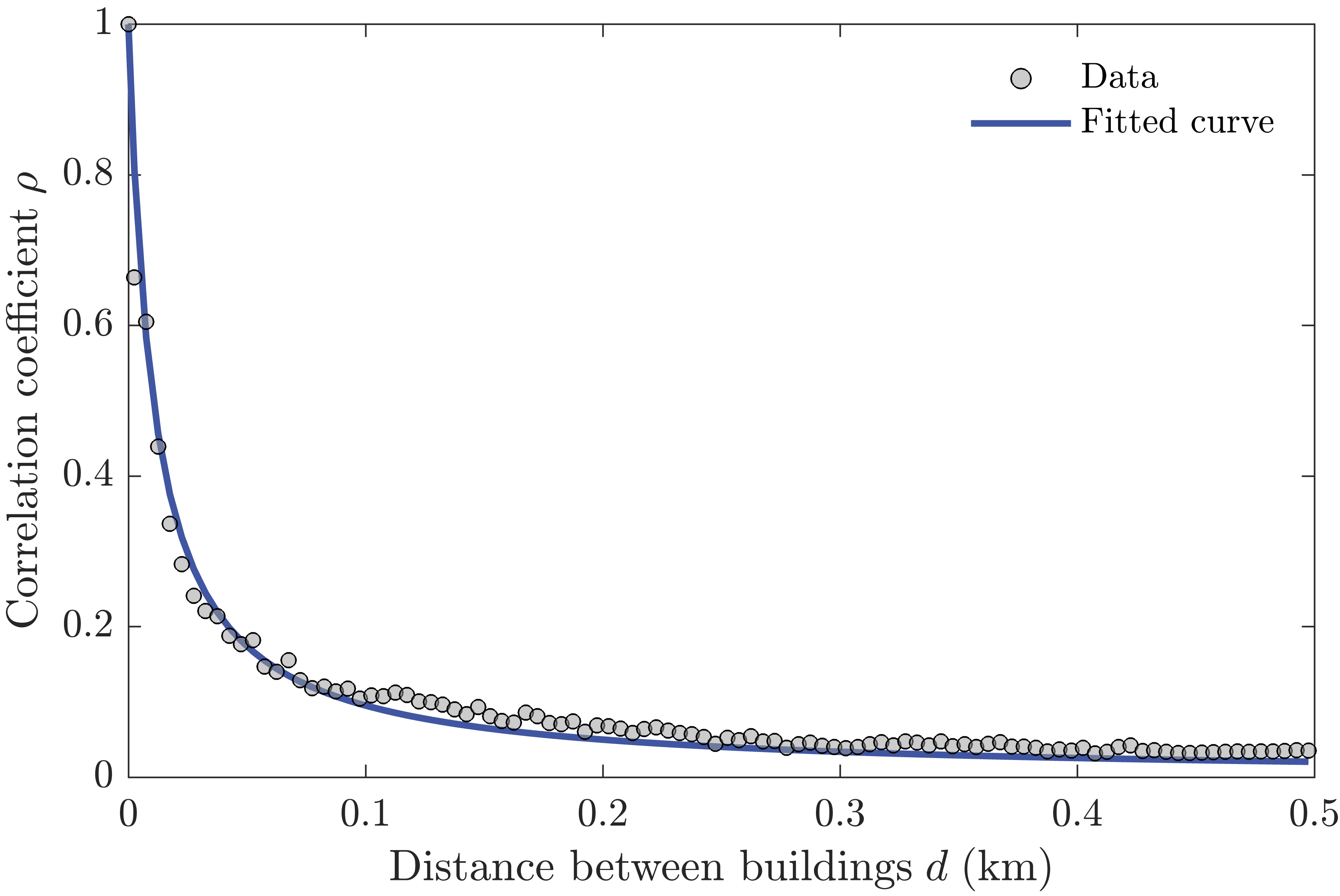}
    \caption{\textbf{The correlation coefficient function $\rho(d)$ fitted to the spatially averaged correlation coefficient values calculated at each distance bin}.}
    \label{Fig:Turkey_correlation_fn}
\end{figure}

\begin{figure}[H]
\begin{subfigure}[t]{.49\textwidth}
    \centering
    \includegraphics[width=1.0\linewidth,trim={0 1.5cm 0 2cm},clip]{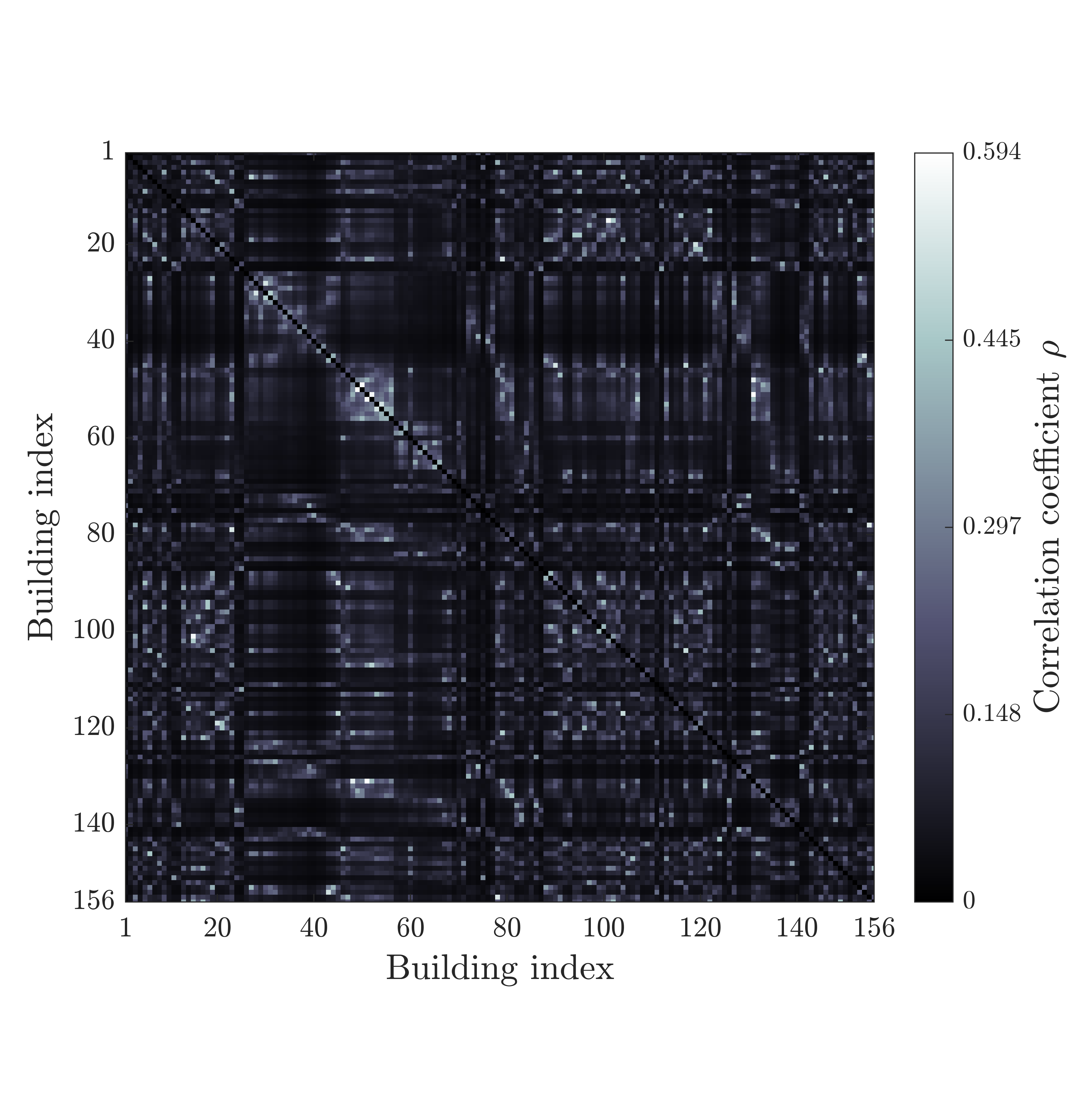}
    \caption{\textbf{}}
    \label{Fig:Turkey_correlation_data}
\end{subfigure}%
\begin{subfigure}[t]{.49\textwidth}
    \centering
    \includegraphics[width=1.0\linewidth,trim={0 1.5cm 0 2cm},clip]{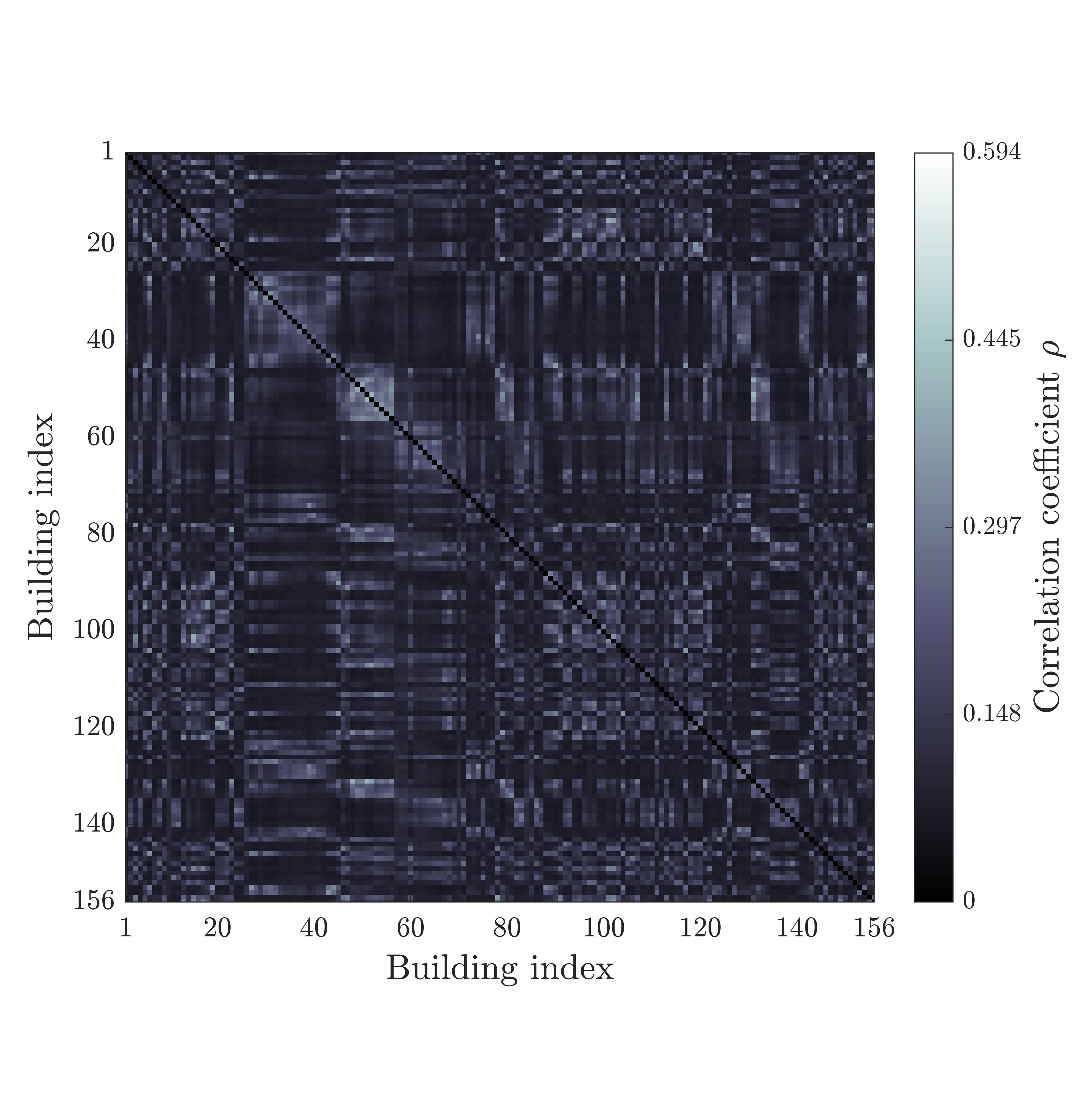}
    \caption{\textbf{}}
    \label{Fig:Turkey_correlation_ising}
\end{subfigure}
\caption{\textbf{Correlation coefficients between each building pair in the target region: (a) calculated from the observation, and (b) reproduced using the Ising model}. The two panels exhibit similar patterns. To enhance the visualization of pairwise correlations, the diagonal terms are excluded.}
\label{Fig:Turkey_correlation}
\end{figure}

To construct the Ising model, we use the following information: the total number of failed buildings, which is 40 out of 156, and the correlation coefficients between each building pair. Given that we have only a single observation under the earthquake event, we assume spatial ergodicity to obtain the correlation coefficients. Specifically, a correlation coefficient function $\rho(d)$, where $d$ is the distance between two buildings, is fitted to the correlation coefficient values obtained by spatially averaging the damage data \textcolor{black}{for $29{,}689$ buildings across the entire Antakya}. The regression equation is found to be $\rho(d)={1}/{\left(1+94.9073d\right)}$, which closely matches the data, as illustrated in Figure \ref{Fig:Turkey_correlation_fn}. \textcolor{black}{Finally, the correlation coefficient values for the $156$ buildings in the target region are calculated using $\rho(d)$, as depicted in Figure~\ref{Fig:Turkey_correlation_data}.}

\begin{figure}[H]
\begin{subfigure}[t]{.48\textwidth}
    \centering
    \includegraphics[height=6.5cm,trim={0 0.8cm 1.2cm 1.2cm},clip]{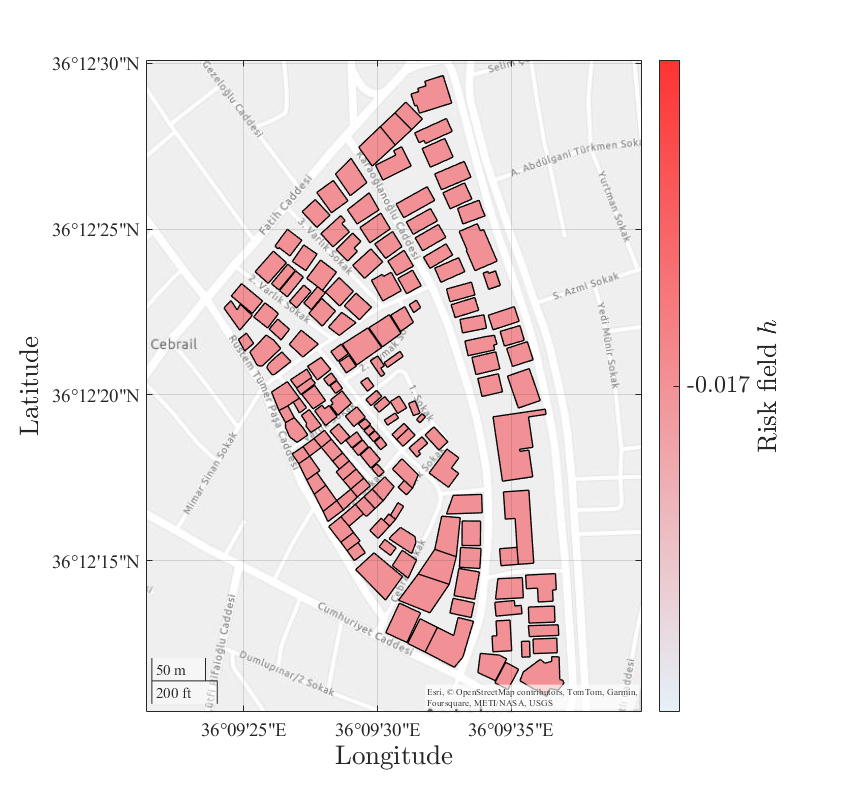}
    \caption{\textbf{}}
    \label{Fig:Turkey_h}
\end{subfigure}%
\begin{subfigure}[t]{.50\textwidth}
    \centering
    \includegraphics[height=6.5cm,trim={0 0.5cm 0 1cm},clip]{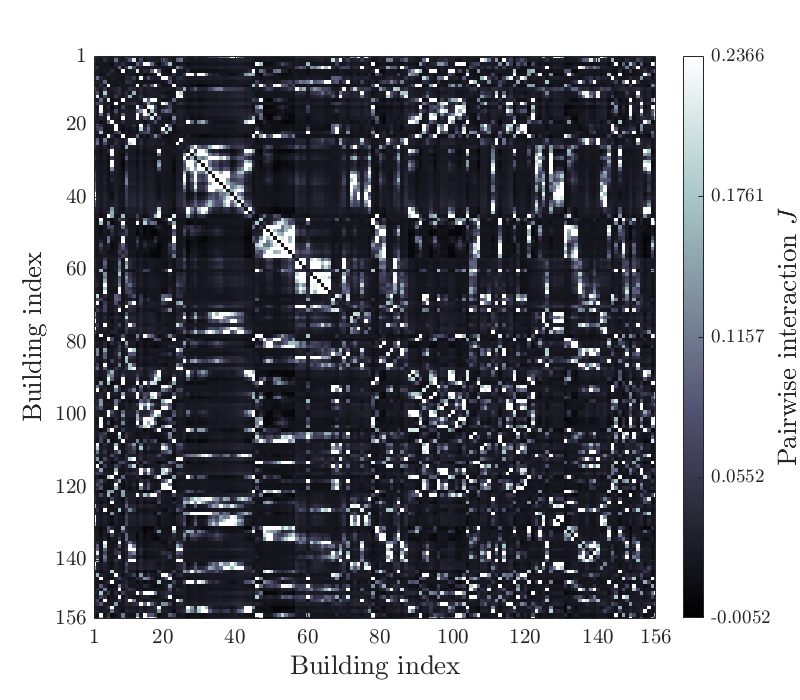}
    \caption{\textbf{}}
    \label{Fig:Turkey_J}
\end{subfigure}
\caption{\textbf{The estimated Ising model parameters for Antakya, Turkey, following the 2023 Turkey-Syria earthquake: (a) risk field $h$ and (b) pairwise interaction $\vect J$}. The prior information \textcolor{black}{only contains the total number of failed buildings instead of individual failure probabilities; thus, a single $h=-0.017$ representing the global risk field for the target region is obtained}. On the other hand, $\vect{J}$ includes interactions for all pairs of buildings, enabled by the correlation function $\rho(d)$.}
\label{Fig:Turkey_param_est}
\end{figure}

The Ising model is then constructed by Boltzmann machine learning. Figure \ref{Fig:Turkey_param_est} illustrates the estimated Ising model parameters. Note that we have only one $h$ representing the global risk field for the target region, rather than a fine-grained $h_i$ for each building. This is consistent with the limited prior information---the total number of failed buildings.

The constructed Ising model reproduces the expected number of failed buildings as 39, which is close to the observed 40. The validity of the constructed Ising model in terms of reproducing the input information can also be seen in Figure \ref{Fig:Turkey_correlation}. The correlation coefficients reproduced using the Ising model, shown as Figure \ref{Fig:Turkey_correlation_ising}, are close to the given values presented in Figure \ref{Fig:Turkey_correlation_data}. \textcolor{black}{Their similarity is quantified by a correlation coefficient computed from flattening each matrix into a vector, and the result is $0.98$}. It is also noteworthy that the pattern of pairwise interactions closely resembles that of the given correlation coefficients. This is because, conditional on the limited input information, the correlation coefficients between damage of buildings are primarily driven by pairwise interactions rather than the global risk field.

\begin{figure}[H]
    \centering
    \includegraphics[width=0.4\linewidth,trim={0 0.8cm 1.2cm 1.2cm},clip]{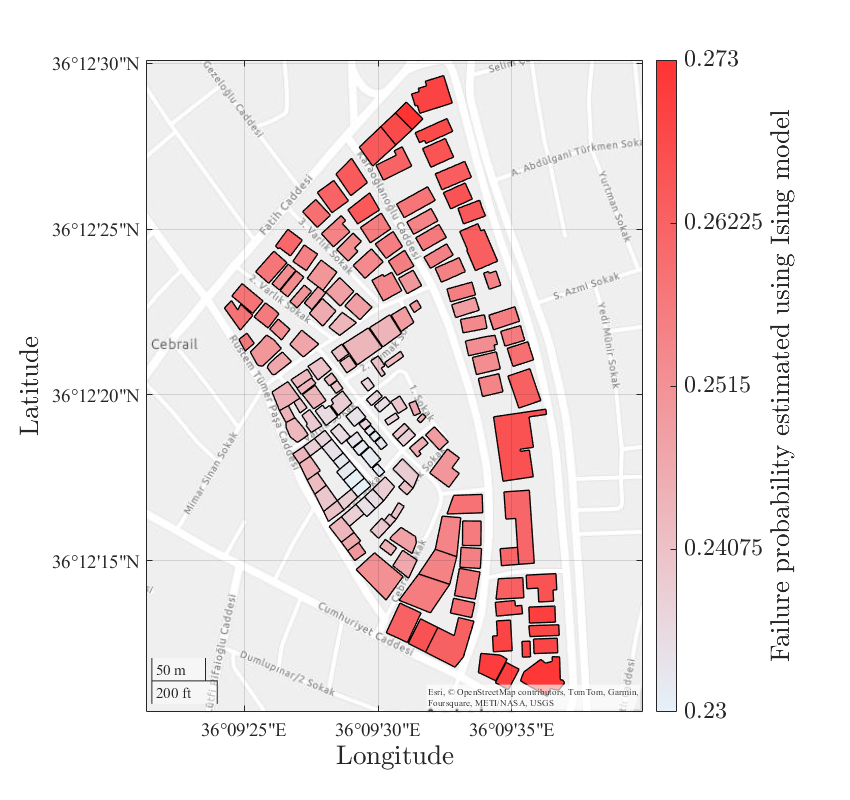}
    \caption{\textbf{Failure probability of each building predicted by the Ising model}. This prediction is based on the current state of knowledge.}
    \label{Fig:Turkey_mean_estimated}
\end{figure}

The Ising model represents a joint probability distribution for all modeled structures over a region, providing a complete statistical description of damage states. Using the model, a total of $10^7$ samples is generated. The failure probability for each building is then obtained, as illustrated in Figure \ref{Fig:Turkey_mean_estimated}. Figure \ref{Fig:Turkey_mean_estimated} shows that closely spaced buildings generally exhibit \textcolor{black}{lower failure probabilities, consistent with the real observation in Figure~\ref{Fig:Turkey_observation}, where about 19\% (14 out of 72) of the buildings failed in the inner area while about 31\% (26 out of 84) failed in the outer area. This result is consistent with the findings of Figure~\ref{Fig:Correspondence_mean}. When the risk field is negative, strong pairwise interactions between closely spaced buildings create a buffer against failure. Conversely, this trend reverses when the risk field is positive.}

\begin{figure}[h]
    \centering
    \includegraphics[width=0.6\linewidth]{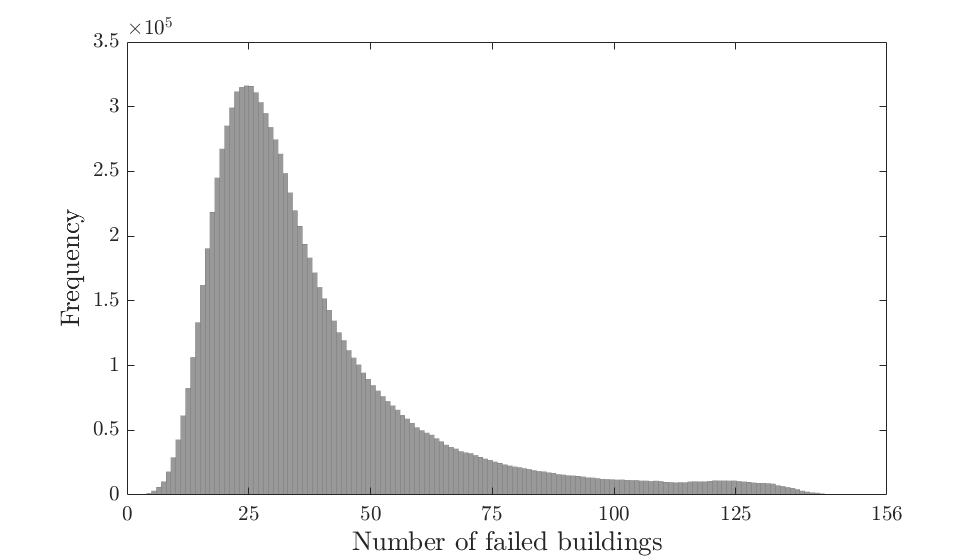}
    \caption{\textbf{Histogram of the number of failed buildings predicted by the Ising model}. The distribution features a long right tail with a small, yet noticeable, second mode.}
    \label{Fig:Turkey_MeanFieldDistribution}
\end{figure}

Figure \ref{Fig:Turkey_MeanFieldDistribution} illustrates a histogram of the number of failed buildings from the $10^7$ samples. It is observed that the mode (most probable number of failed buildings) is around 25, although the expected number is 39. Moreover, the probability of more than 100 buildings will fail is 3.85\%, which is not negligible. This is because the distribution of the number of failed buildings has a long right tail. Indeed, it is well known that the Ising model exhibits \textcolor{black}{a bimodal distribution} for the global mean-field (number of failed buildings in this example) when pairwise interactions are large \cite{Kadanoff2000StatisticalRenormalization}. Although bimodality is only noticeable in Figure \ref{Fig:Turkey_MeanFieldDistribution}, in regions with higher correlation coefficients, such as residential zones with similar houses, the bimodality would be more pronounced and a collective failure would be non-negligible.

\textcolor{black}{The inferences drawn from Figures \ref{Fig:Turkey_mean_estimated} and \ref{Fig:Turkey_MeanFieldDistribution} exemplify how the Ising model can be used to organize raw data and draw inferences from it. It is important to note that these inferences are conditional on the current state of knowledge; they may not reflect the \textit{truth} associated with an idealized state of sufficient knowledge. However, inferences from the Ising model can be improved as higher-fidelity information about the target region becomes available.}

\subsection{Example 2: Ising model based on fragility functions}
\noindent
In this example, the Ising model is constructed from samples generated by the R2D tool provided by the NHERI SimCenter. A neighborhood in Pacific Heights, San Francisco, is chosen as the modeled region, containing 182 buildings. A earthquake scenario of magnitude 7.0 is assumed. The epicenter is on the San Andreas Fault, at latitude 37.9, longitude -122.432, with a depth of 80 kilometers. Given the earthquake scenario, the R2D computes peak ground accelerations (PGA) at each building site using the ground motion prediction equation (GMPE) developed in \citet{Boore2014NGA-West2Earthquakes}. The models from \citet{Baker2008CorrelationModels} and \citet{Markhvida2017EffectModel} are used to account for the inter- and intra-event spatial correlations in PGAs, respectively. Given the PGA values at each building site, the damage state for each building is evaluated using fragility curves provided by Hazus \cite{FEMA2014Multi-hazardManual}, leveraging structural information integrated into the R2D tool. This includes geographical location, plan area, year built, number of stories, and structural type of the buildings. Due to limitations in the current R2D tool, correlations between the capacities of buildings are not considered. Subsequently, 10,000 samples are generated, each sample being a vector representing the damage states for the 182 buildings.

\begin{figure}[H]
\begin{subfigure}[t]{.49\textwidth}
    \centering
    \includegraphics[width=1.0\linewidth,trim={0 0.8cm 0 1.5cm},clip]{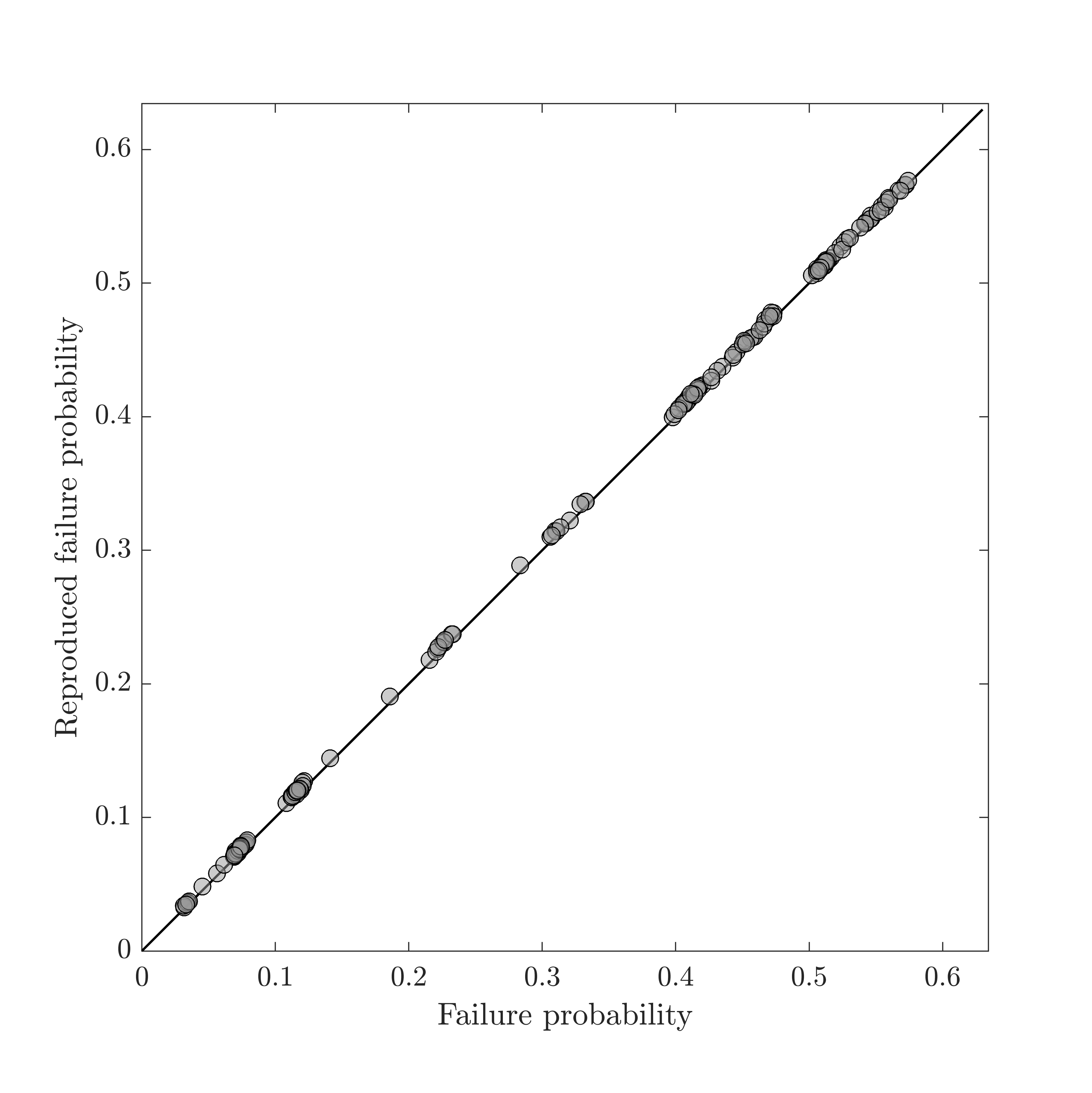}
    \caption{\textbf{}}
    \label{Fig:R2D_M70_Pf_scatter}
\end{subfigure}%
\begin{subfigure}[t]{.49\textwidth}
    \centering
    \includegraphics[width=1.0\linewidth,trim={0 0.8cm 0 1.5cm},clip]{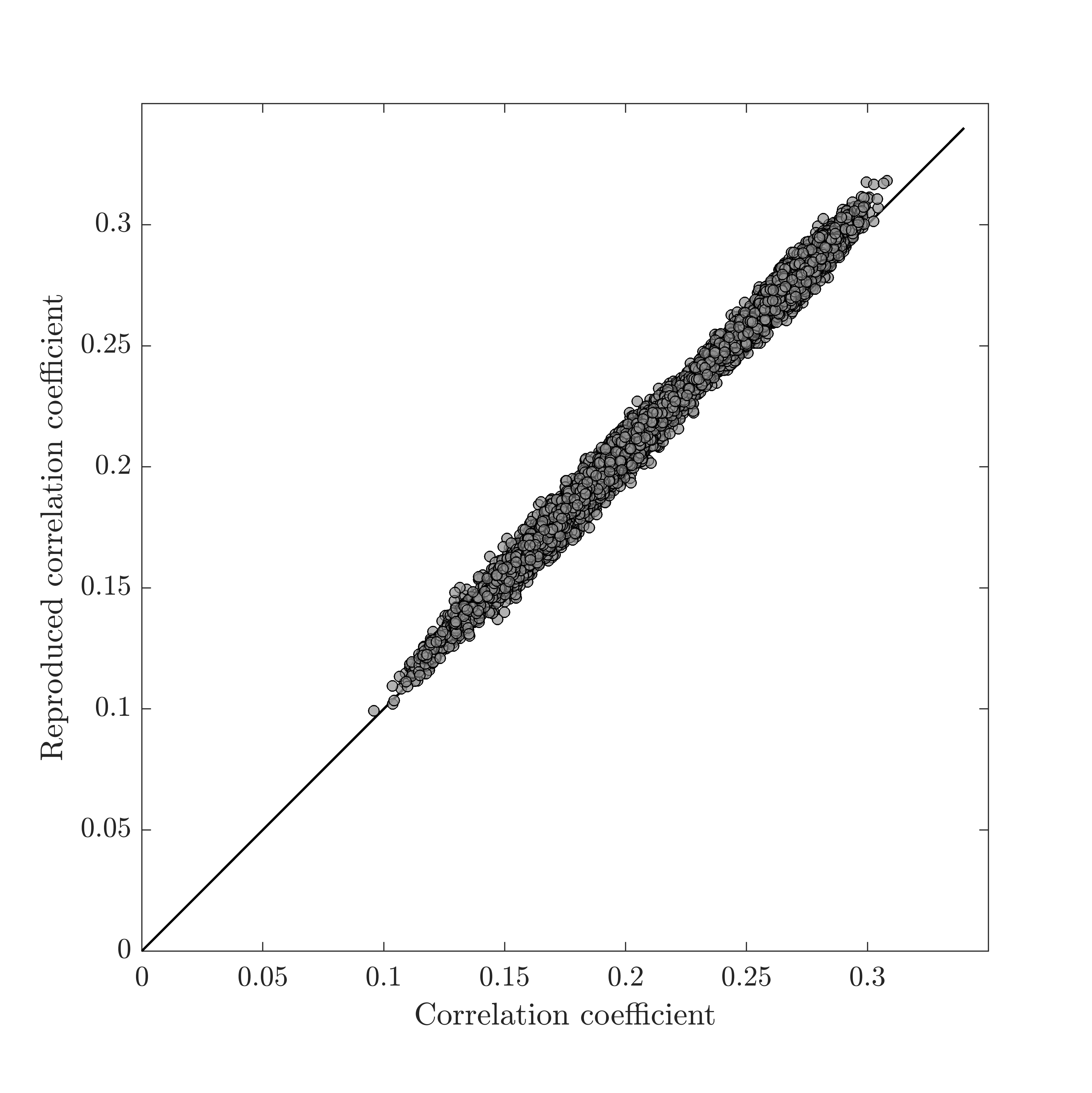}
    \caption{\textbf{}}
    \label{Fig:R2D_M70_corr_scatter}
\end{subfigure}
\caption{\textbf{Comparison between the given statistics and those reproduced by the Ising model: (a) failure probability, and (b) correlation coefficient}. The solid diagonal line signifies perfect reconstruction.}
\label{Fig:R2D_M70_scatter}
\end{figure}

Using the samples, we obtain the failure probability for each building and the correlation coefficient matrix of the damage states. The Ising model parameters are then estimated. Figure \ref{Fig:R2D_M70_scatter} shows a comparison between the given statistics and those reproduced by the Ising model. It is observed that the Ising model accurately reproduces the given statistics. 

\begin{figure}[H]
\begin{subfigure}[t]{.49\textwidth}
    \centering
    \includegraphics[width=1.0\textwidth]{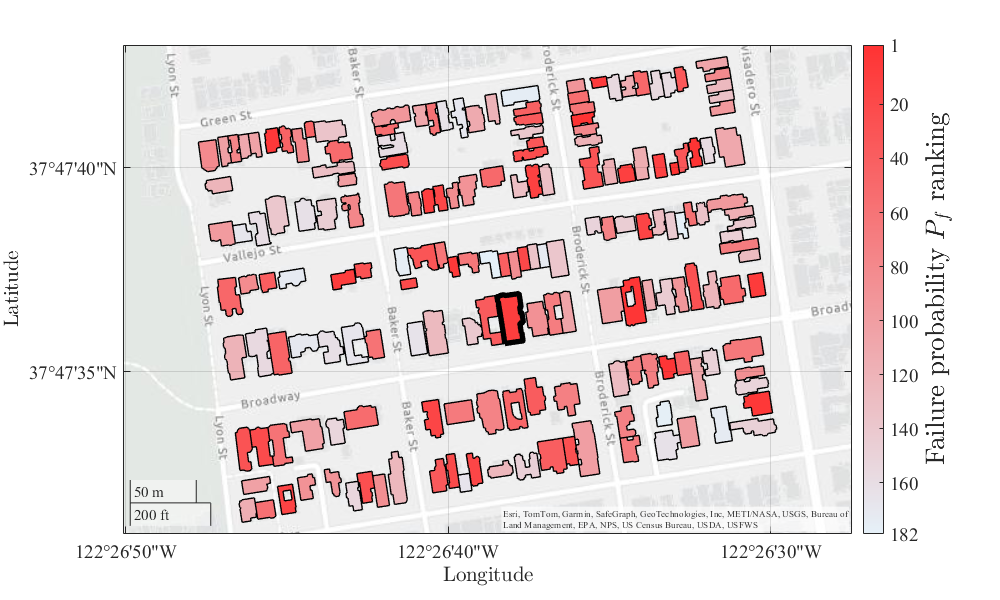}
    \caption{\textbf{}}
    \label{Fig:R2D_M70_Pf}
\end{subfigure}
\begin{subfigure}[t]{.49\textwidth}
    \centering
    \includegraphics[width=1.0\textwidth]{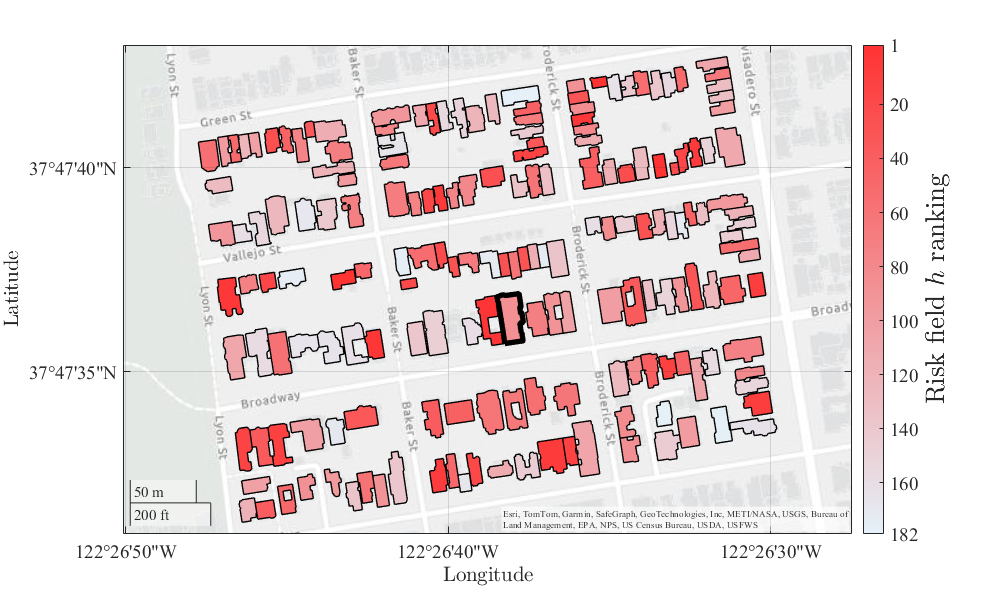}
    \caption{\textbf{}}
    \label{Fig:R2D_M70_h}
\end{subfigure}
\caption{\textbf{Illustrations of (a) the failure probabilities and (b) the risk field $\vect h$ values for each building in Pacific Heights, San Francisco}. Building \#108, which is of particular importance, is highlighted by a bold black solid line.}
\label{Fig:R2D_M70}
\end{figure}

\begin{figure}[H]
    \centering
        \includegraphics[scale=0.52]
    {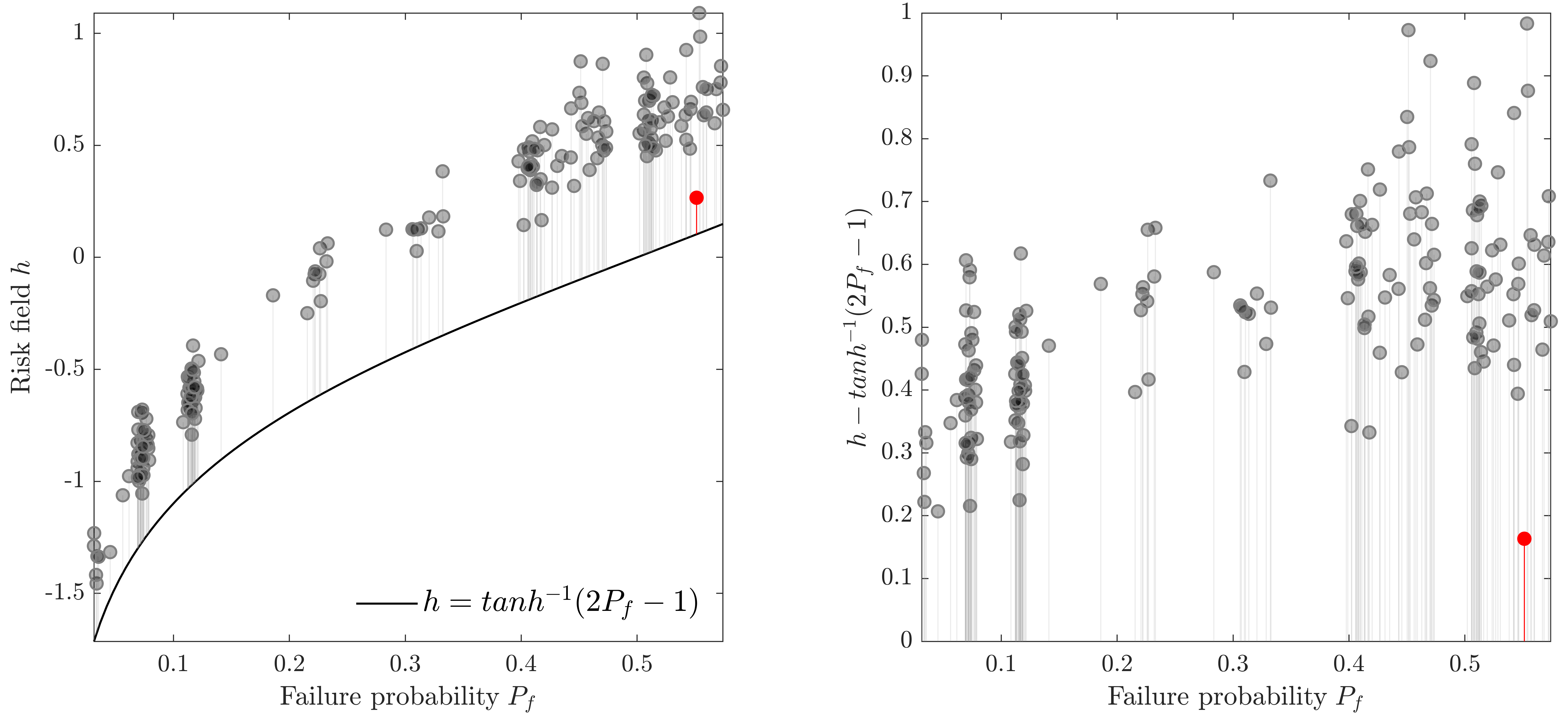}
    \caption{\textcolor{black}{\textbf{Comparison between the risk fields and the failure probabilities}. The black solid curve in the left panel represents the exact relationship between $h$ and $P_f$ when the damage states are independent, as a special case of the mean-field approximation represented by Eq.~\eqref{Eq:meanfield_inverse}. The distance from each point to the reference curve of independence is visualized as a solid thin line, which is highlighted in the right panel. Building \#108, marked in red, is identified as an outlier because, compared to other buildings with similar failure probabilities, its distance to the reference curve is distinctly small.}}
    \label{Fig:R2D_M70_Pf_h_comparison}
\end{figure}

\begin{figure}[H]
    \centering
    \includegraphics[width=1.0\textwidth,trim={1cm 0 1cm 0},clip]{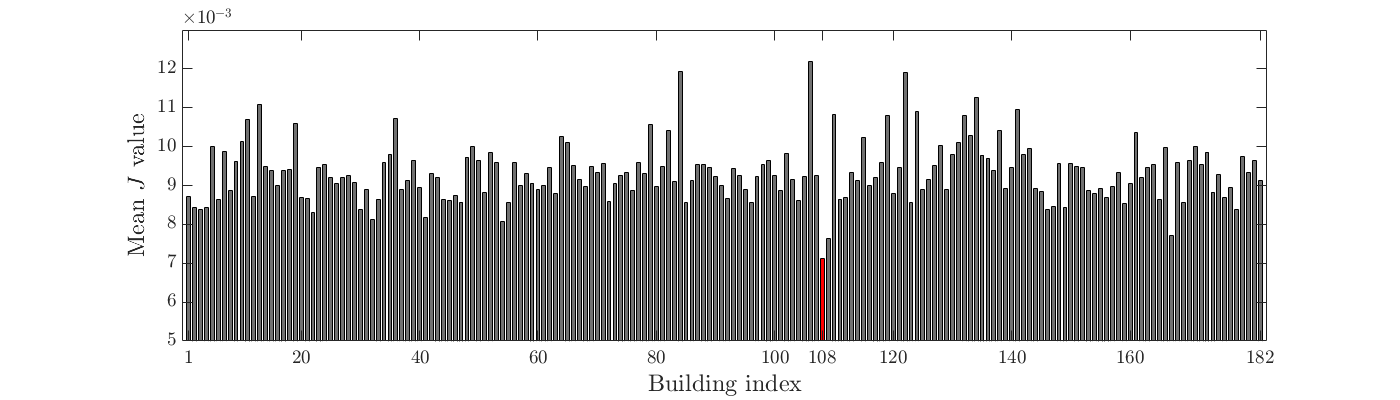}
    \caption{\textbf{The average pairwise interaction value for each building}. For the $i$-th building, we average its interaction with other buildings to obtain $\sum_{j\neq i} J_{ij}/\left(n-1\right)$. The average interaction for building \#108 is highlighted in red.}
    \label{Fig:R2D_M70_meanJ_bar}
\end{figure}

Figure \ref{Fig:R2D_M70} compares the failure probabilities provided by the R2D tool with the risk field values derived from the Ising model, with the color bars \textcolor{black}{indicating the rankings of buildings based on these two quantities, where $1$ denotes the highest and $182$ the lowest. The spatial patterns are overall similar, but differences are noticeable. These differences are highlighted in Figure \ref{Fig:R2D_M70_Pf_h_comparison} by comparing the risk fields to the failure probabilities, with the reference curve indicating independence between damage states. If all damage states are independent, the points should be distributed along the reference curve and the rankings from the two metrics will be identical. Additionally, if all points follow a monotonically increasing trend, the rankings are identical. In Figure \ref{Fig:R2D_M70_Pf_h_comparison}, however, the risk field value varies significantly even for similar failure probabilities. This multiplicity is due to the interactions among structures. For instance, the building indexed as $108$ stands out in Figure \ref{Fig:R2D_M70_Pf_h_comparison} as being distinctly close to the reference curve, implying that it is \textit{closest} to independence. In fact, Figure \ref{Fig:R2D_M70_meanJ_bar} shows that building \#108 deviates from the rest by having the smallest average pairwise interaction value. This suggests that the failure event of building \#108 has a minimal effect on the state of the other buildings.} These observations are supported by the mean-field approximation in Eq.~\eqref{Eq:meanfield_inverse}, where the discrepancy between $h_i$ and $m_i$ is contributed by the interaction terms. \textbf{Therefore, contrasting the risk field $\vect{h}$ with individual failure probabilities reveals the dominance of collective behavior among structures}. \textcolor{black}{The collective behavior is realized by the term ${\vect x}\tr\vect J\vect x$ in the Ising model, which directly alters the \textit{joint distribution}; the effect reaches beyond second-order statistics.} 
    
\begin{figure}[H]
    \centering
    \includegraphics[width=1\textwidth]{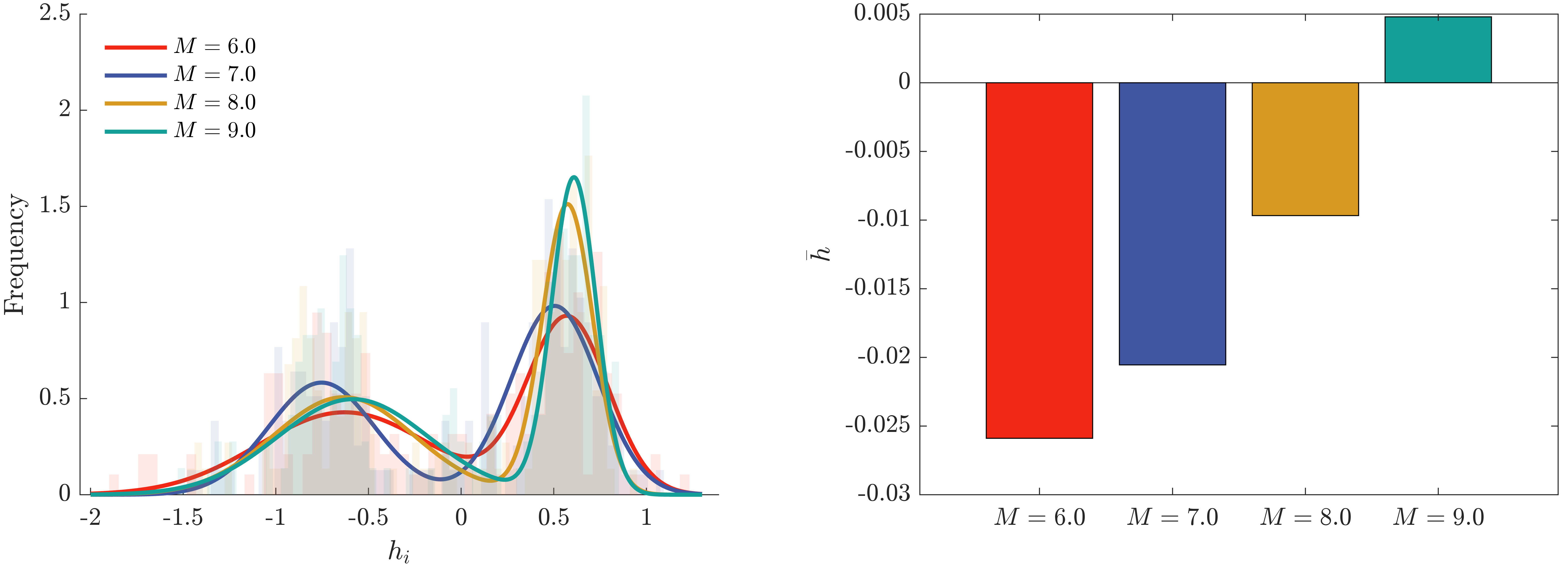}
    \caption{\textbf{The relationship between the risk field $\vect h$ and earthquake magnitude}. The left panel shows the histogram of $h_i$ for four earthquake magnitudes, while the right panel illustrates the spatial mean of $h_i$, $\bar{h} = \sum_i h_i / n$. It is observed that the positive mode of $h_i$ becomes more pronounced with increasing earthquake magnitude. Additionally, the spatial mean $\bar{h}$, which is proportional to the probability of all structures collectively failing, monotonically increases with magnitude.}
    \label{Fig:R2D_merge_Pf_h}
\end{figure}

\begin{figure}[H]
    \centering
    \includegraphics[width=1.0\textwidth]{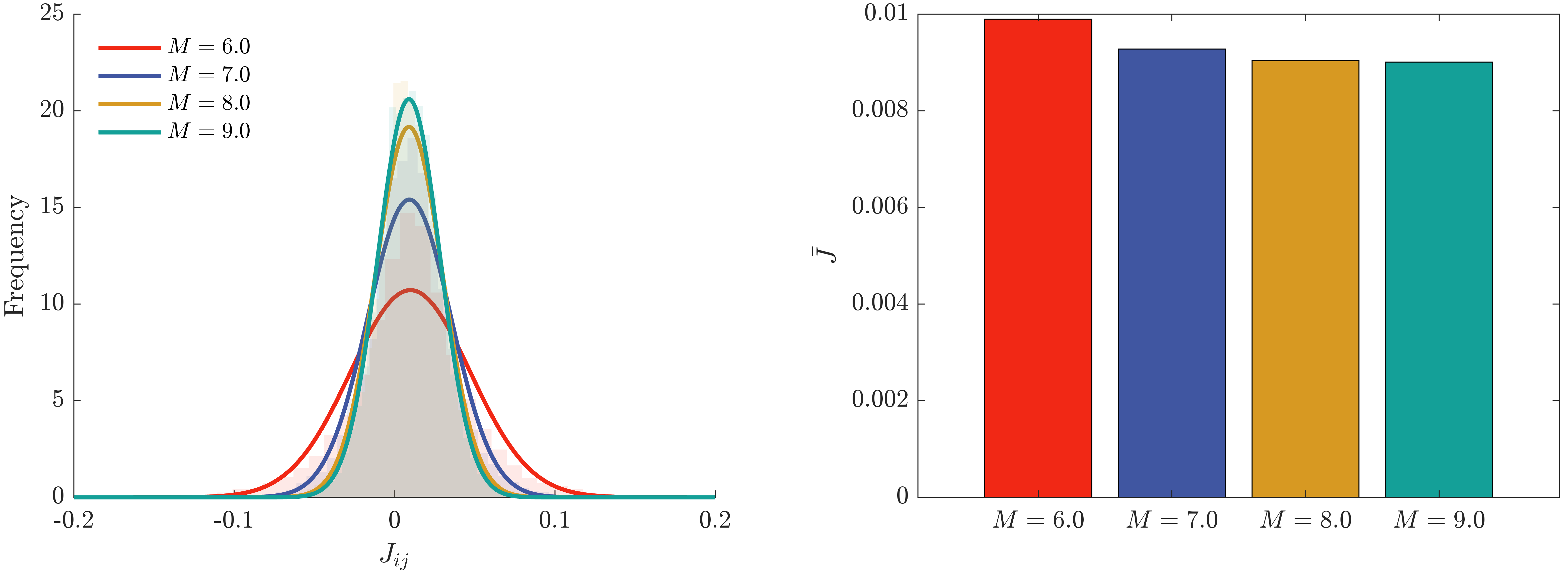}
    \caption{\textbf{The relationship between the pairwise interaction $\vect J$ and earthquake magnitude}. The left panel shows the histogram of $J_{ij}$ for four earthquake magnitudes, while the right panel illustrates the spatial mean of $J_{ij}$, $\bar{J} = \sum_{i\neq j} J_{ij} / \left(n-1\right)^2$. It is observed that $J_{ij}$ becomes more concentrated with increasing earthquake magnitude, while $\bar{J}$ decreases slowly with magnitude.}
    \label{Fig:R2D_merge_J_histogram}
\end{figure}

Lastly, we examine the relationship between the risk field and earthquake magnitudes. Beyond the existing scenario of magnitude $7.0$, we construct Ising models for earthquake magnitudes of $6.0$, $8.0$, and $9.0$. The results, presented in Figure \ref{Fig:R2D_merge_Pf_h}, show that the right/positive mode of $h_i$ becomes more dominant as the magnitude increases. The left/negative mode of $h_i$ persists across all magnitudes. This is because the mean failure probability is around $0.5$ even at a magnitude of $9.0$, suggesting a significant inertia toward remaining safe, i.e., negative $h_i$, among many structures. The relationship between the pairwise interaction and earthquake magnitudes is illustrated by Figure \ref{Fig:R2D_merge_J_histogram}. The results suggest that the dispersion of $J_{ij}$ decreases as the magnitude increases, but the spatial mean of $J_{ij}$ across all pairs is not sensitive to the magnitude. This weak dependency between $\vect{J}$ and magnitude may be attributed to the fact that $\vect{J}$ primarily reflects the inherent interdependencies among structures.

\section{Conclusions} \label{Section:Conclusions}
\noindent
This study introduces the long-range Ising model as a joint probability distribution of damage states of structures for regional seismic analysis. Using the principle of maximum entropy, the long-range Ising model is derived under the constraints of the first- and second-order cross-moments of the damage states. The Ising model parameters, $\vect h$ and $\vect J$, are interpreted as a risk field acting at each structure and the pairwise interaction between structures, respectively. Through simple demonstrative examples, the complex interplay between the Ising model parameters and the mean and correlation coefficient of damage states is illustrated. It is found that the mean damage state of a structure is primarily influenced by the risk field acting on it, although the pairwise interactions also play a role in either sharpening or diminishing this effect. Similarly, the correlation coefficient for the damage states between a pair of structures is influenced by all pairwise interactions involving those structures, with the magnitude of the risk field either amplifying or mitigating this effect. The engineering relevance of the Ising model is further illustrated by two simulation test cases. The first example considers the damage states of buildings in Antakya, Turkey, following the 2023 Turkey-Syria earthquake, the second example studies neighborhood in Pacific Heights, San Francisco, under sythetic earthquake scenarios generated by the Regional Resilience Determination (R2D) tool of the NHERI SimCenter. It is found that for both examples the Ising model can reproduce the given information. More importantly, the Ising model offers insights on the collective behavior of structural failures. Specifically, it is found that the discrepancy between the risk field values and the individual failure probabilities signifies the dominance of collective behavior in group failures or non-failures. 
    
The Ising model serves as a fundamental component for more advanced statistical mechanics methods to analyze the multi-scale behaviors, phase transitions, and scaling laws in complex multi-body systems. This work establishes a basis for further extensions, suggesting several promising research directions:
\begin{itemize}
    \item Development of \textit{Renormalization Group} \cite{Kadanoff2000StatisticalRenormalization} method for modeling and understanding the multi-scale behavior and phase transitions in regional seismic responses.
    \item Early-warning systems for system-level phase transitions and criticality.
    \item Ising models for multiple damage states and functionalities of networked civil infrastructures.
    \item {\color{black} Comprehensive comparative studies using high-fidelity data on regional seismic responses. }
    \item Development of efficient computational methods for estimating Ising model parameters.
\end{itemize}
    
\appendix
\section{Derivation of the Ising model using the principle of maximum entropy} \label{Appendix:Derivation}
\noindent
The goal is to find the joint probability $P_I$ that meets the given constraints and also maximizes the entropy. Applying Lagrange multipliers for the constraints described in Eqs.~\eqref{Eq:1storder} and \eqref{Eq:2ndorder}, maximizing the entropy $S$ is equivalent to maximizing:
    \begin{equation} \label{Eq_appendix:Lagrange}
    \begin{aligned}
    L=&-\sum_{I=1}^N P_I \log{P_I}-\alpha\left(1-\sum_{I=1}^N P_I\right)\\
    &-\sum_{i=1}^n h_i\left(\mathcal{M}_{1,i}-\sum_{I=1}^N x_{iI}P_I\right)-\sum_{i>j} J_{ij}\left(\mathcal{M}_{2,ij}-\sum_{I=1}^N x_{iI}x_{jI}P_I\right)\,,    
    \end{aligned}
    \end{equation}
where $\alpha$, $h_i$, and $J_{ij}$ are Lagrange multipliers. Taking the partial derivative of Eq.~\eqref{Eq_appendix:Lagrange} with respect to $P_I$, we obtain:
    \begin{equation} \label{Eq_appendix:Lagrange_derivative}
    \frac{\partial L}{\partial P_I}=-\log{P_I}-1+\alpha+\sum_{i=1}^n h_ix_{iI}+\sum_{i>j} J_{ij}x_{iI}x_{jI}\,.
    \end{equation}
Finally, we obtain $P_I$ that meets $\frac{\partial S}{\partial P_I}=0$ as follows:
\begin{equation}
    \begin{aligned}
       P_I &= \exp{\left(-1+\alpha+\sum_{i=1}^n h_ix_{iI}+\sum_{i>j} J_{ij}x_{iI}x_{jI}\right)} \nonumber \\
            &= \frac{1}{Z}\exp{\left(\sum_{i=1}^n h_ix_{iI}+\sum_{i>j} J_{ij}x_{iI}x_{jI}\right)} \,,
    \end{aligned}
\end{equation}
where $Z=\exp{\left(1-\alpha\right)}$.

\section{Derivation of the mean-field approximation} \label{Appendix:Meanfieldsolution}
\noindent
\textcolor{black}{The mean-field approximation is foundational in statistical mechanics, and its derivation can be found in many papers and textbooks \cite{Kadanoff2000StatisticalRenormalization, Ma2019ModernPhenomena, Utermohlen2018MeanModel, Nguyen2017InverseScience}. However, for completeness, we present a derivation in the context of the long-range Ising model.}

The mean-field approximation assumes that the fluctuation in each component is relatively small compared with its mean-field \cite{Utermohlen2018MeanModel}. Then, $x_ix_j$ can be approximated by:
\begin{equation}
    \begin{aligned}
       x_ix_j &= \left(m_i+\left(x_i-m_i\right)\right)\left(m_j+{\color{black}{\left(x_j-m_j\right)}}\right) \\
           &\approx m_im_j+m_i\left(x_j-m_j\right)+m_j\left(x_i-m_i\right) \\
           &= m_ix_j+m_jx_i-m_im_j \,. 
    \end{aligned}
\end{equation}
Subsequently, the normalizing constant $Z$ can be \textcolor{black}{obtained by summing the unnormalized probabilities in Eq.~\eqref{Eq:Isingmodel} over all configurations $I$,} expressed by:
\begin{equation} \label{Eq:meanfield_Z}
    \begin{aligned}
        Z &= \sum_I\exp{\left(\sum_{i=1}^{n}h_ix_{i}+\sum_{i>j}J_{ij}x_{i}x_{j}\right)} \\
          &\approx \sum_I\exp{\left(\sum_{i=1}^{n}h_ix_{i}+\sum_{i>j}J_{ij}m_{i}x_{j}+\sum_{i>j}J_{ij}m_{j}x_{i}-\sum_{i>j}J_{ij}m_{i}m_{j}\right)} \\
          &= \sum_I\exp{\left(\sum_{i=1}^{n}h_ix_{i}+\sum_{i\neq j}J_{ij}m_{j}x_{i}-\frac{1}{2}\sum_{i\neq j}J_{ij}m_{i}m_{j}\right)} \\
          &= \sum_I\exp{\left(\sum_{i=1}^{n}\Tilde{h}_ix_{i}-\frac{1}{2}\sum_{i\neq j}J_{ij}m_{i}m_{j}\right)}. \\
          &= c \sum_I\exp{\left(\sum_{i=1}^{n}\Tilde{h}_ix_{i}\right)} \,,
    \end{aligned}
\end{equation}
where $\Tilde{h}_i = h_i+\sum_{j\neq i}J_{ij}m_{j}$ is the \textit{effective external field} that incorporates the effect of pairwise interactions, and $c\equiv\exp\left(-\frac{1}{2}\sum_{i\neq j}J_{ij}m_{i}m_{j}\right)$ is introduced for brevity. The symmetry of $\vect J$, $J_{ij}=J_{ji}$, was used for the third line of \textcolor{black}{Eq.~\eqref{Eq:meanfield_Z}}. We further simplify $Z$ into:
\begin{equation}
    \begin{aligned}
        Z &= c \sum_I \prod_{i=1}^{n} \exp{\Tilde{h}_i x_{i}} \\
          &= {\color{black}c \sum_I \left(\exp{\Tilde{h}_1 x_{1}}\right)\left(\exp{\Tilde{h}_2 x_{2}}\right)\cdots\left(\exp{\Tilde{h}_n x_{n}}\right)} \\
          &= {\color{black}c \left(\exp\left(\Tilde{h}_1\right)+\exp\left(-\Tilde{h}_1\right)\right)\left(\exp\left(\Tilde{h}_2\right)+\exp\left(-\Tilde{h}_2\right)\right)\ldots\left(\exp\left(\Tilde{h}_n\right)+\exp\left(-\Tilde{h}_n\right)\right)} \\
          &= c \prod_{i=1}^{n} \left(\exp\left(\Tilde{h}_i\right)+\exp\left(-\Tilde{h}_i\right) \right)\\
          &= c \prod_{i=1}^{n} 2\cosh{\Tilde{h}_i}\,,
    \end{aligned}
\end{equation}
which is followed by: 
\begin{equation}
\ln{Z} = \ln c+n\ln{2}+\sum_{i=1}^n \ln{\cosh{\Tilde{h}_i}}\,.
\end{equation}
Finally, using the properties of $m_i=\frac{\partial \ln{Z}}{\partial h_i}$ and $c_{ij}=\frac{\partial m_i}{\partial h_j}$, the self-consistent equations for $m_i$ and $c_{ij}$ are derived as: 
\begin{equation}
    \begin{aligned}
     m_i &= \frac{\partial \ln{Z}}{\partial h_i}\\
            &= \tanh\left(\Tilde{h}_i\right) \\
            &= \tanh\left({h_i+\sum_{j\neq i}J_{ij}m_{j}}\right)\,, \\
        c_{ij} &= \frac{\partial m_i}{\partial h_j} \\
               &= \left(1-m_i^2\right)\left(\delta_{ij}+\sum_{k\neq i}J_{ik}c_{kj}\right)\,. 
    \end{aligned}
\end{equation}

\bibliography{Reference}

\end{document}